\documentclass[acmsmall, nonacm]{acmart}
\usepackage{amsmath,amsfonts}
\usepackage{algorithm}
\usepackage{algpseudocode}
\usepackage{array}
\usepackage{textcomp}
\usepackage{stfloats}
\usepackage{url}
\usepackage{verbatim}
\usepackage{graphicx}
\usepackage{cite}
\usepackage{multirow}
\usepackage[dvipsnames]{xcolor}
\usepackage{bbding}
\usepackage{colortbl}
\usepackage{enumitem}
\usepackage{wrapfig}
\usepackage{makecell}
\usepackage{booktabs}
\usepackage{multirow}
\usepackage{amsmath}
\usepackage{booktabs}
\usepackage{subcaption}
\usepackage[skins,breakable]{tcolorbox}
\usepackage[table]{xcolor}

\authorsaddresses{}

\begin{document}

\title{Knowledge Graph-based Retrieval-Augmented Generation for Schema Matching}

\author{Chuangtao Ma}
\affiliation{%
  \institution{Aalborg University}
  \city{Aalborg}
  \country{Denmark}
}
\email{chuma@cs.aau.dk}

\author{Sriom Chakrabarti}
\affiliation{%
  \institution{Aalborg University}
  \city{Aalborg}
  \country{Denmark}
}
\email{scha@cs.aau.dk}

\author{Arijit Khan}
\affiliation{%
  \institution{Bowling Green State University}
  \city{Ohio}
  \country{USA}
}
\affiliation{%
  \institution{Aalborg University}
  \city{Aalborg}
  \country{Denmark}
}
\email{arijitk@bgsu.edu}

\author{Bálint Molnár}
\affiliation{%
  \institution{Eötvös Loránd University}
  \city{Budapest}
  \country{Hungary}
}
\email{molnarba@inf.elte.hu}


\begin{abstract}
  Traditional similarity-based schema matching methods fail to resolve semantic ambiguities and conflicts in complex, domain-specific mappings due to a lack of commonsense and domain-specific knowledge. In parallel, large language models (LLMs) without sufficient background knowledge tend to hallucinate, further complicating schema matching. To overcome these limitations, we introduce KG-RAG4SM, a Knowledge Graph-based Retrieval-Augmented Generation for Schema Matching. This method integrates novel retrieval mechanisms, including LLM-based, vector-based, graph traversal-based, and hybrid techniques with advanced reranking schemes to extract the most relevant subgraphs from large external KGs. Our approach achieves superior results without requiring re-training, as evidenced by improvements of 39.79\% in precision and 30.56\% in \texttt{F1} score over LLM-based state-of-the-art (SOTA) methods (e.g., Jellyfish-8B) on the MIMIC dataset, and enhancements of 69.20\% in precision and 21.99\% in \texttt{F1} score over pre-trained language model (PLM)-based SOTA methods (e.g., SMAT) on the Synthea dataset. Additionally, KG-RAG4SM scales effectively across KGs from commonsense KGs to domain-specific ones, while its instruction-tuning knowledge exclusion component ensures truthful performance in end-to-end schema matching tasks despite the suboptimal relevance of KG context. Our empirical analysis and case studies from real-world scenarios show that the hallucination problem in LLMs for schema matching is well mitigated by our solution.
\end{abstract}

\maketitle

\section{Introduction}

Schema matching reconciles multiple database schemas by identifying semantic correspondences between source and target schema elements through mapping techniques~\citep{hai2007schema}. In conventional approaches, similarity scores are computed for each pair of attribute names and accepted if they exceed a predefined threshold. However, these methods often lack the depth to recognize implicit semantic relationships, such as synonyms, abbreviations, and different terminologies from heterogeneous schemas, due to their inherent reliance on surface-level similarity. This semantic heterogeneity frequently necessitates human intervention and domain expertise to validate mapping decisions~\citep{Rahm2011}. Given the vast number of entities and attributes across heterogeneous databases, such manual oversight is impractical. For example, disparate management information systems (MIS), including electronic health records (EHR), must often be integrated into a standardized database to ensure efficient data access and management. Here is an example of schema matching between the e-MedSolution data model~\citep{ma2020knowledge} and the OMOP CDM data model~\citep{jiang2017consensus} from the real-world scenario.
\begin{figure}[htbp]
\centering
\includegraphics[width=0.65\textwidth]{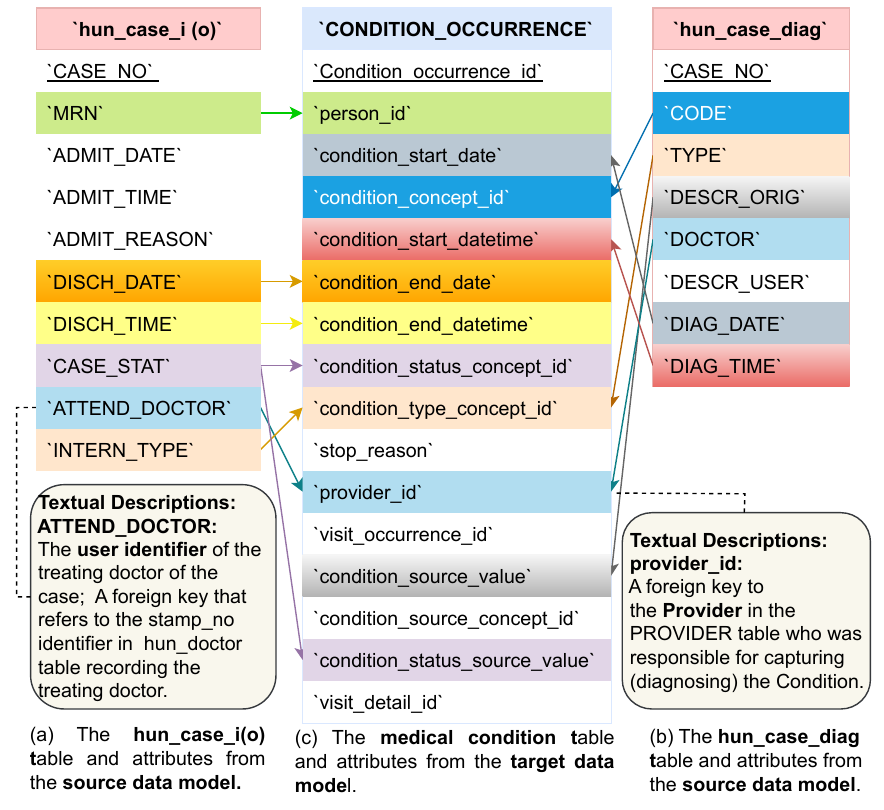}
\caption{Example of schema matching in the EHR data model. Table (a) and Table (b) are from the source schema, and Table (c) is from the target schema. The textual description for attributes originates from the database design documents. Solid arrows indicate the corresponding mappings share the same color. }\label{fig:example}
\end{figure}
As shown in Figure \ref{fig:example}, the \texttt{ATTEND\_DOCTOR} from \texttt{hun\_case\_i(o)} table is a semantically similar attribute with the \texttt{DOCTOR} of \texttt{hun\_case\_diag} table in the source schema, and both of them should be matched with the \texttt{provider\_id} from \texttt{CONDITION\_OCCURRENCE} table in the target schema. 
Traditional schema matching approaches struggle with such complex scenarios due to semantic heterogeneity and ambiguity, as they do not effectively capture the semantic information and the meaning of the tables~\citep{freire2025large}.

In recent years, language models have demonstrated superior capabilities in schema matching and heterogeneous data integration. For instance, past works propose to integrate the textual description of entities and attributes with large language models (LLMs)~\citep{Parciak2024}, e.g., fine-tuning LLMs~\citep{zhang2024jellyfish, steiner2024fine}, prompting LLMs~\citep{feng2024cost}, and augmenting LLMs~\citep{sheetrit2024rematch} to generate the top-$k$ mappings.
Other works employ pre-trained language models (PLMs) to classify the potential mappings~\citep{zhang2023schema, tu2023unicorn}.
However, the SOTA LLM-based and PLM-based schema matching is still incapable of handling the conflicts and ambiguities at the semantic level (e.g., acronyms, abbreviations, relatedness, etc.) in certain complex mapping cases, for instance, identifying the matches between \texttt{ATTEND\_DOCTOR} and \texttt{provider\_id} in the above example. 
The reasons are twofold: \textbf{(1)} The schema matching in real-world data integration usually involves identifying the semantic correspondence between attributes spanning across multiple tables from heterogeneous schemas; \textbf{(2)} LLM and PLM-based methods necessitate textual descriptions and background knowledge to analyze the given tables and attribute names; however, the background knowledge is often absent in real-world databases.
Without acquiring the background knowledge, the LLM-based method could only incorporate the textual information at the syntactic level
\citep{zhang2022using, ma2022knowledge}, which limits their capabilities to detect deeper semantic correspondences and find many-to-one mappings with similar semantic meanings~\citep{bodensohn2025unveiling}.

Knowledge Graphs (KGs) are one of the widely used knowledge bases, where the rich semantic knowledge and fact-based relations are structured in the form of $\langle$subject, predicate, object$\rangle$ triples~\citep{cheng2024neural}. 
They store large-scale, real-world facts using graph structures in a highly curated manner~\citep{zhang2021smedbert}, where a node (subject or object) represents an entity, and an edge (predicate) denotes a relationship between two entities. 
The structured semantic knowledge in external commonsense KGs makes it possible to enhance the language models with knowledge inference by incorporating commonsense knowledge, factual knowledge, and domain knowledge, thereby resolving complex matching cases without any re-training of LLMs. 
In particular, commonsense, fact-based, and domain-specific KGs can be introduced to augment the LLMs and better handle the above complex schema matching case. 
This is mainly facilitated by the commonsense and fact-based KGs, such as Wikidata~\citep{vrandevcic2014wikidata}, which contain several common knowledge facts for taxonomic and hierarchical relationships, acronyms, abbreviations, relatedness, etc., which are always accessible and retrievable.
Meanwhile, domain-specific KGs, such as Unified Medical Language System (UMLS) and SNOMED-CT~\citep{snomed}, also contain rich relationships between, e.g., healthcare and medical terms, concepts, and codes with synonyms, which are useful for augmenting the LLMs to handle complex schema matching cases in the healthcare databases. 
    
\begin{figure}[h]
\centering
\includegraphics[width=0.8\textwidth]{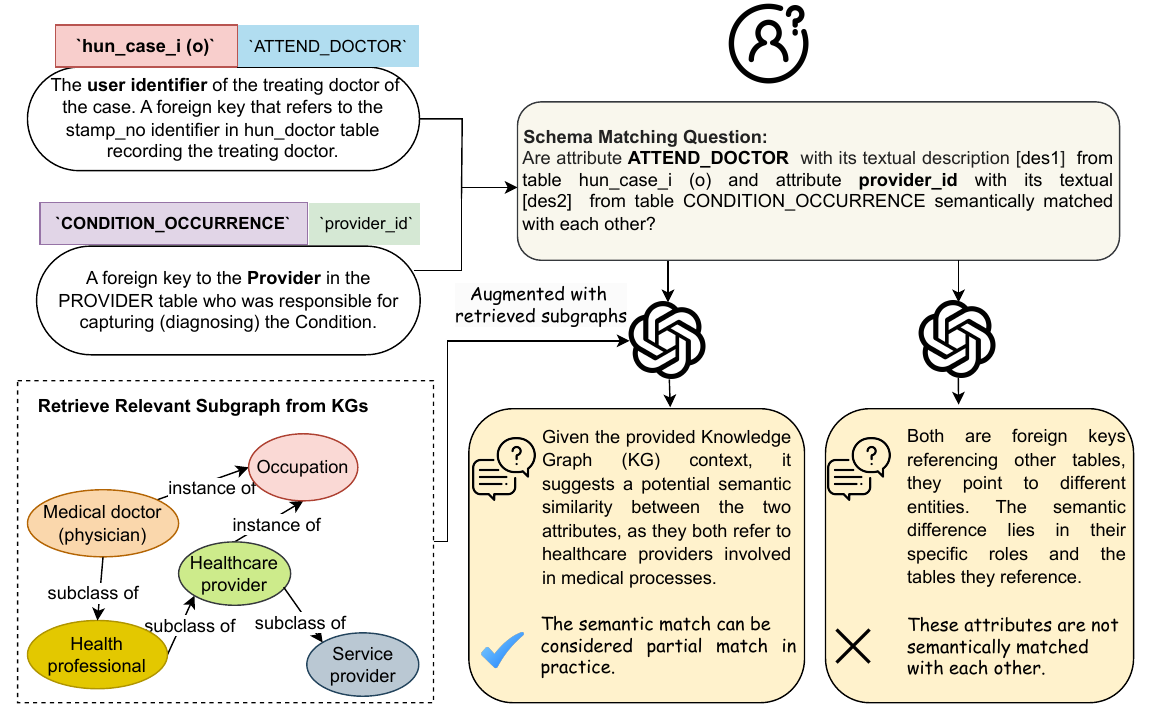}
\caption{The effectiveness of KG context in augmenting LLMs for complex schema matching.}\label{fig:kg_role}
\end{figure}

For instance, consider the mapping between \texttt{ATTEND\_DOCTOR} and \texttt{provider\_id}. The retrieved information from a commonsense KG can augment and guide LLMs to respond with the correct answer. As demonstrated in Figure \ref{fig:kg_role}, the retrieved KG context can help LLMs find the potential semantic similarity between the two attributes, which augments the capability of LLMs for complex schema matching tasks.

Given that the rich fact-based semantic knowledge inherent in KGs has demonstrated remarkable performance in knowledge reasoning and semantic disambiguation~\citep{RenHL20,0001KWJY20}, KGs can be exploited to augment the capability of LLMs for the aforementioned complex schema matching.
Unlike mainstream Graph RAG and KG-RAG for knowledge graph question answering (KGQA), where local, small-scale KGs can be built from benchmark datasets or text chunks~\citep{hu2024grag, sarmah2024hybridrag,sanmartin2024kg, li2025simple}, there are no internal KGs within databases that can be directly leveraged to augment LLMs for complex schema matching because the required semantic knowledge exists in external and large KGs. 
It is not technically feasible to augment the LLMs with these large-scale KGs directly, because the 
contexts of LLMs have a finite length~\citep{Wei2024}.
Besides, context-poisoning issues may occur if too much irrelevant knowledge is utilized to augment the prompt. As a result, the performance of the LLM-based schema matching will decrease~\citep{Parciak2024}.  
Therefore, the key technical challenge of augmenting LLMs with external KGs lies in how to effectively and efficiently retrieve the relevant subgraphs from the large-scale external KGs, for instance, Wikidata includes billions of triples~\citep{vrandevcic2014wikidata}.

To address the above challenges, we propose a \textbf{K}nowledge \textbf{G}raph-based \textbf{R}etrieval \textbf{A}ugmented \textbf{G}eneration for \textbf{S}chema \textbf{M}atching (KG-RAG4SM). 
Our method integrates vector-based, LLM-based, and graph traversal-based retrievals to identify the relevant subgraphs from large and external KGs and further prunes the irrelevant knowledge by using a ranking scheme. Then, it leverages the retrieved subgraphs to augment the LLMs and prompts for generating the final results for the complex schema-matching task. 

\begin{figure}[h]
\centering
\includegraphics[width=0.8\textwidth]{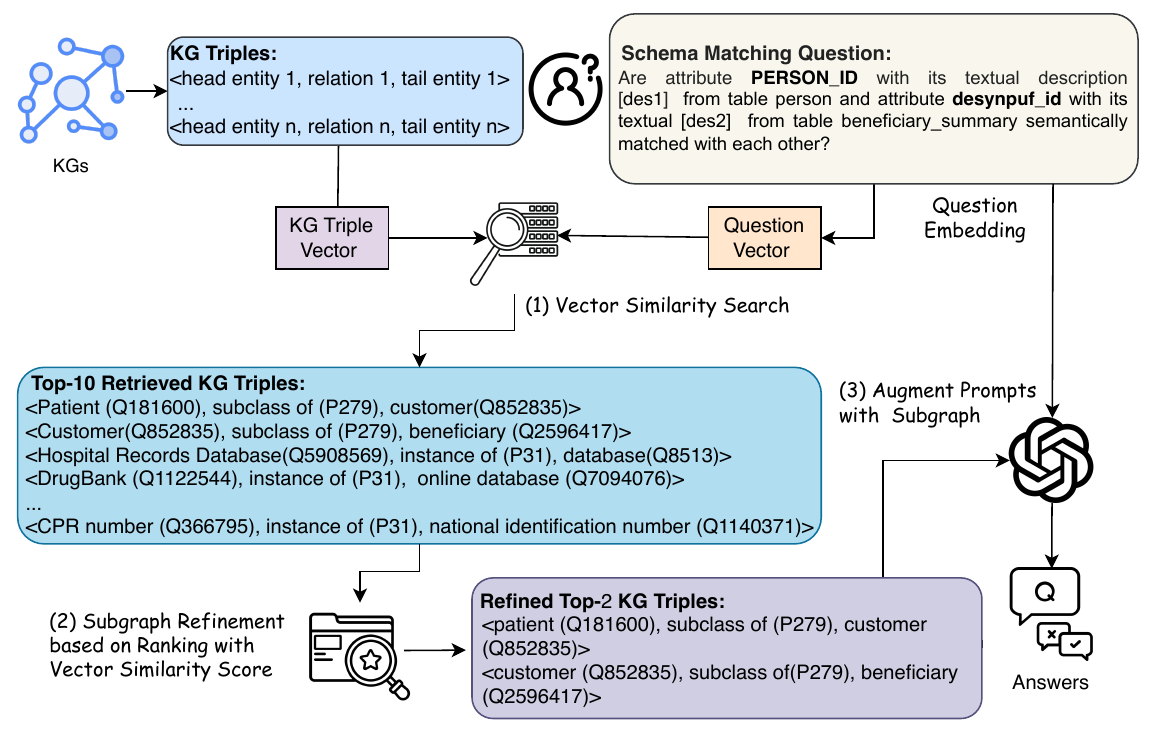}
\caption{Overview of our proposed KG-RAG4SM method. Given schema matching questions and an external knowledge graph (KG), we \textbf{(1)} retrieve the relevant KG triples based on vector similarity between question embeddings and KG triple embeddings; \textbf{(2)} prune the retrieved relevant KG triples with vector similarity-based rankings; \textbf{(3)} augment prompts with the retrieved and refined subgraphs and generate the final answer for the given schema matching questions.}\label{fig:kgrag4sm_method}
\end{figure}

As shown in Figure \ref{fig:kgrag4sm_method}, we introduce novel knowledge graph retrieval and subgraph pruning techniques to retrieve and prune the relevant subgraphs from large KGs for augmenting the prompts and LLMs.
Our method compares favorably with baseline models
\citep{zhang2021smat,tu2023unicorn,zhang2024jellyfish} in handling the semantic conflicts of complex mapping cases, by retrieving the relevant subgraphs from large and external KGs and reconciling them with LLMs for augmenting the schema matching.

To the best of our knowledge, {\em we are the first to augment the LLMs and prompts for schema matching tasks with the retrieved relevant subgraphs from large and external KGs}. 
Unlike the emerging Graph RAG~\citep{hu2024grag, sarmah2024hybridrag, WangFZLM26} and KG-RAG~\citep{sanmartin2024kg, li2025simple} that retrieve local neighbors or subgraphs from small, self-constructed graphs out of text chunks, or the inherent KGs derived from the knowledge graph question answering (KGQA) dataset, our KG-RAG4SM retrieves relevant subgraphs directly from large external commonsense KGs. This enables the system to leverage rich factual knowledge to resolve complex cross-domain schema matching tasks.
Mainstream graph reasoning methods~\citep{buehler2025situ, LuoLHP24, SunXTW0GNSG24, zhao2024kg} generate multi‑step paths through a KG, typically via beam search or chain‑of‑thought prompting, which introduces significant computational overhead on large graphs and complex queries. KG‑RAG4SM takes a different route: it uses hybrid retrieval, knowledge refinement, and exclusion mechanisms to extract a compact, task‑relevant factual subgraph from large external KGs, enabling one‑shot grounding rather than iterative path construction. By grounding the LLM in retrieved precise facts instead of internalizing graph thinking patterns, KG‑RAG4SM delivers more efficient and accurate cross‑domain schema matching performance.
Our contributions are outlined as follows. 
\begin{itemize}[noitemsep, leftmargin=0.3cm, topsep=0pt]
\item We present a knowledge graph-based retrieval-augmented generation for schema matching that retrieves and integrates the relevant subgraphs from large external KGs into LLM-based schema matching via in-context learning. 
\item We propose novel subgraph retrieval techniques utilizing LLM-based, vector-based, and graph traversal-based approaches, as well as a hybrid method that integrates these strategies to extract pertinent subgraphs from a large external KG and further enhance LLM-based schema matching.
\item We prune the retrieved subgraphs by removing the irrelevant entities and edges based on a ranking scheme to provide concise and factual knowledge for augmenting the LLMs and address the context poisoning issue in schema matching.
\item We conduct a thorough experimental evaluation on six benchmark datasets across domains and a case study from the real-world healthcare database to demonstrate the effectiveness, efficiency, and scalability across KG domains of our solutions.
\item  We introduce \textsf{EMED}, a new schema matching dataset from the real-world healthcare domain. This will promote future research on domain-specific schema matching and data integration.
\end{itemize}

The remainder of the paper is organized as follows. The preliminaries and the problem formulation are given in \S  \ref{sec2}. The details of our proposed KG-RAG4SM method are introduced in \S \ref{sec3}. Then, experimental evaluation and the case study are conducted in \S \ref{sec4}, and the related work is summarized in \S \ref{sec5}. Finally, the conclusion and the future work are stated in \S \ref{sec6}. 
The additional empirical evaluation, ablation study, sensitivity analysis, and complexity analysis are provided in the Appendix.  

\section{Preliminaries and Problem Formulation}
\label{sec2}
We first discuss preliminaries and then state the problem of knowledge-graph-based retrieval-augmented generation for schema matching.

\subsection{Preliminaries}
We start with preliminaries on LLMs and schema matching, subgraph retrieval, and knowledge graph-based retrieval-augmented generation for schema matching. Table \ref{tab1} presents the frequently used notations in this work.

\begin{table}
    \caption{Frequently Used Notations.}
    \label{tab1}
    \centering
    \small
    \begin{tabular}{c|l} 
    \toprule
         Notation & Description \\
         \midrule
       $\mathcal{KG}, \mathcal{G}, P, P^{\prime}$ & The large knowledge graph and its subgraph, paths, refined and ranked paths \\ \hline
       $\left\langle h, r, t \right\rangle$ & The triple is composed of head entity $h$, relation $r$, and tail entity $t$ \\ \hline
       $E, R$ & The retrieved list of the entities and relations from large-size $\mathcal{KG}$ \\ \hline
       $S_s$, $S_t$ & The source schema and target schema \\ \hline
       $T_s$, $T_t$, $tab_s$, $tab_t$ & The tables and an individual table of the source schema and target schema \\ \hline 
       $A_s$, $A_t$, $att_s$, $att_t$ & The attributes and an individual attribute of the source schema and target schema \\ \hline
       $\mathcal{C}_{mp}, c_{mp}$ & The set of candidate matching pairs and a candidate matching pair  \\ \hline
       $\theta$ & The schema matching task is modeled with binary classification  \\ \hline
       $q, a, \mathcal{A}$ & The query or question that is fed to LLMs to generate the answer $a$ and answer set $\mathcal{A}$ \\ 
       \bottomrule
    \end{tabular}
\end{table}

\subsubsection{LLMs as a Schema Matcher}
Schema matching aims to reconcile the semantic ambiguity between source and target schema by establishing semantic correspondences between source and target schema.
Let a source database schema $S_s$ with tables $T_s$ and attributes $A_s$ be represented as a set of attribute-table pairs $(att_s, tab_s) \in S_s$, and let a target database schema $S_t$ with tables $T_t$ and attributes $A_t$ consist of attribute-table pairs $(att_t, tab_t) \in S_t$. The schema matching task is to find correspondences between $(att_s, tab_s) \in S_s$ and $(att_t, tab_t) \in S_t$.
The set of candidate matching pairs $\mathcal{C}_{mp}$ is generated via the Cartesian product of $S_s$ and $S_t$:
$$\mathcal{C}_{mp} = S_s \times S_t = \{ ((att_s, tab_s), (att_t, tab_t)) \mid (att_s, tab_s) \in S_s, (att_t, tab_t) \in S_t \}$$
In this setting, the schema matching problem is treated as a classification problem, which can be solved by learning a binary classifier $\theta$:
$$\text{LLM}(\theta)=\{c_{mp} | c_{mp} \in \mathcal{C}_{mp}\} \rightarrow \{\text{Yes, No}\} $$
where each attribute correspondence is an element of the set of candidate matching pairs $\mathcal{C}_{mp}$.
It is worth mentioning that the method with the above setting allows LLMs to handle not only the one-to-one mapping but also the one-to-many, many-to-one, and many-to-many mappings.

\subsubsection{Graph RAG}
Graph retrieval-augmented generation (Graph RAG)~\citep{peng2024graph} usually retrieves knowledge from graphs to enhance the capability of LLMs to generate answers for the given task.
Due to the context poisoning of LLMs, large-scale graphs are not technically suitable for augmenting LLMs directly. 
Subgraph retrieval or path selection~\citep{liu2024-knowledge-graph} aims to retrieve the relevant subgraph or path from the large graph based on vector similarity and ranking.

Given a question $q$, a set of candidate answers $\mathcal{A}$, and a large-scale graph $G$ employed to augment the LLM, the goal is to generate the optimal answer $a^*$:
$$a^* = \arg \max_{a \in \mathcal{A}} p(a \mid q, G)$$
where $p(a \mid q, G)$ is modeled via a subgraph retriever $p(\mathcal{G} \mid q, G)$ and an LLM-based answer generator $p(a \mid q, \mathcal{G})$. 
Specifically, the subgraph retriever aims to extract the top-$k$ subgraphs $\mathcal{G} \subset G$ or paths $P$ that are most relevant for the LLM to answer the question $q$.

\subsection{Problem Statement}
We formally define the problem of knowledge graph-based retrieval-augmented generation for schema matching.
Given a large language model (LLM), an external and large-scale knowledge graph $\mathcal{KG}$, a retrieved subgraph $\mathcal{G}$, and a schema matching task $\theta$, formally, the KG-RAG4SM is a transformation $\mathcal{F}$:
$$\text{LLM}^{\prime}:=\mathcal{F}\left(\text{LLM}, \mathcal{G}\right) \quad \text{s.t.} \quad a_{\text{LLM}^{\prime}(\theta)}>a_{\text{LLM}(\theta)}$$
where $\text{LLM}^{\prime}$ denotes the LLMs augmented with the retrieved subgraph $\mathcal{G}$ ($\mathcal{G} \subset \mathcal{KG}$) from large-size $\mathcal{KG}$.
The constraint $a_{\text{LLM}^{\prime}(\theta)}>a_{\text{LLM}(\theta)}$ describes that the overall correctness of the generated answers for schema matching by LLMs augmented with retrieved subgraphs is better than the answers generated by LLMs without retrieved subgraphs from the KG for schema matching.

In our KG-RAG4SM, KG-RAG is introduced to augment the generation phase of LLMs for complex tasks by integrating LLM-based, vector-based, and graph traversal-based, as well as hybrid retrievals to retrieve the relevant subgraphs from large KGs.
Technically, KG-RAG augments the question using the top-$k$ subgraphs from $\mathcal{KG}$ based on vector similarity and ranking.

\section{Methodology}
\label{sec3}
We provide an overview and technical details of the proposed solutions. 
\textit{Our key technical contributions include a thorough investigation of the novel KG-RAG paradigm for the schema matching problem by designing efficient subgraph retrieval methods via LLM-based, vector-based, and graph traversal-based retrievals, as well as a hybrid approach that integrates them after analyzing their trade-offs, followed by effective ranking schemes to provide precise knowledge for augmenting the LLMs in schema matching.}

\subsection{Leveraging LLM as a Schema Matcher} \label{sec3.1}
In the pipeline of LLM as a schema matcher, a binary question $q$ for each matching pair is generated based on the source attribute name $att_s$, target attribute name $att_t$, and their textual descriptions from the source schema $S_s$ and target schema $S_t$.
Then the question $q$ generated based on each matching pair $c_{mp}$, including few-shot examples with their correct answers and explanations encapsulated with the prompt, is fed into the LLMs to generate the correct answer. 
A prompt including few-shot examples with their correct answers and explanations is given in Appendix~\ref{app:prompt_llms_as_schema_matcher}, where the placeholder \{question\} will be replaced with the real questions from the selected dataset. Other placeholders [d1] and [d2] are descriptions for attribute1 and attribute2, respectively. Together with Example 2, they are omitted due to brevity. 

\subsection{Augmenting LLM with Knowledge Graphs} \label{sec3.2}
We empirically find that the attribute names and their textual descriptions alone are not sufficient in the LLM schema matching pipeline for complex matching scenarios that require semantic contextual information. To extract semantic similarity effectively and efficiently in the form of relevant subgraphs from external knowledge graphs (KGs), we develop a number of retrieval methods as described below. They generally follow three main steps: {\bf (1)} retrieval of relevant entities and relations from the KG w.r.t. the question; {\bf (2)} then identifying relevant subgraphs induced by the relevant entities from the KG; and finally {\bf(3)} pruning and ranking of the relevant subgraphs to provide more precise semantic contextual information in LLM prompts.

\subsubsection{Retrieving Relevant Entities from the KG} \label{subs:entityretrieval} Before retrieving the subgraph from KG, we need to retrieve the entities from large-size KGs that are relevant to the given question of schema matching. We follow one of the two developed paradigms for this purpose. 

\noindent \textbf{LLM-based Entity Retrieval.} Given a schema-matching question, in this approach, we ask an LLM (e.g., GPT-4o-mini) to retrieve a few relevant entities from the KG. Our prompts for this purpose, including a few-shot examples with their correct answers and explanations, are illustrated in Appendix~\ref {app:prompts_llms_as_entity_retriever}. The placeholders \{question\} are replaced with the real questions from our selected dataset. Nevertheless, LLMs can hallucinate; thus, they may report some entities that do not, in fact, exist in the commonsense KG. Therefore, we need to verify the retrieved entities by LLMs before retrieving the subgraphs among these entities.
It is worth mentioning that LLM-based entity retrieval only works well for commonsense and fact-based KGs (e.g., Wikidata) rather than domain-specific KGs, since the former KGs or their open text data forms (e.g., Wikipedia) might have already been utilized to pre-train the LLMs. As a result, these KGs are well-known to LLMs, while the domain-specific KGs may not be well-known to LLMs.

\noindent \textbf{Vector-similarity based Entity Retrieval.} 
This method leverages vector similarity search to retrieve the relevant entities and relations from the KG that are relevant to the schema matching questions. 
By representing both the questions and the knowledge graph components as dense vector embeddings, the vector search techniques are used to identify the top-$k$ relevant entities based on efficient computation of similarity scores. This approach ensures high-quality and scalable retrieval of knowledge graph components, making it well-suited for applications in schema matching and semantic search.

\noindent \underline{\textit{Offline KG embedding and vector storage.}} The embeddings for KG entities and relations are generated in an offline manner using pre-trained transformer-based language models such as RoBERTa~\citep{liu2019roberta}, as it demonstrated strong results for vector search in terms of both effectiveness and computational efficiency~\citep{monir2024vectorsearch}.
These embeddings, along with associated metadata (e.g., entity or relation identifiers), are stored in a vector database such as ChromaDB~\citep{PanWL24}. The storage process includes indexing within the vector database to facilitate efficient retrieval of relevant embeddings during online querying.

\noindent \underline{\textit{Online question embedding.}} The online component starts by generating embeddings for each question. 
In particular, question embeddings are generated using Sentence-BERT~\citep{reimers2019sentence}, which employs the same pre-trained model, RoBERTa, and is specifically designed for sentence embeddings because it is good at capturing the semantic meaning of sentences. 
This approach works for all KGs, including domain-specific ones, because sentence transformers, such as RoBERTa, are pre-trained on diverse datasets, enabling them to capture general semantic patterns and relationships. By embedding entities and relationships into a universal vector space, they ensure that semantic similarity is preserved across domains.

\noindent \underline{\textit{Online vector search.}} 
To retrieve the top-$k$ most similar entities and relationships, we employ cosine similarity as the distance metric. 
Cosine similarity is chosen after extensive experimentation against alternative distance metrics such as Euclidean and Manhattan distances. 
Although Euclidean and Manhattan distances are effective for capturing absolute differences in vector space, cosine similarity is superior in our context as it measures the angular alignment between vectors, disregarding their magnitude. 
We experimentally observe that cosine similarity consistently outperforms other metrics in our vector-based search and retrieval. 
Cosine similarity measures the cosine of the angle between two vectors, and it is defined as:
$$ \text{Cosine Similarity}(A, B) = \frac{A \cdot B}{\|A\| \|B\|} \nonumber $$
Here, $A$ and $B$ represent the embeddings of the question and a KG entity or relationship, respectively, $\cdot$ denotes the dot product, and $\|\,\|$ represents the magnitude of a vector.

Additionally, we use the HNSW (Hierarchical Navigable Small World) indexes implemented in the vector databases for efficient vector retrieval~\citep{malkov2020efficient}. 
The HNSW index constructs a graph-based structure where nodes represent data points, and edges connect neighboring points based on similarity, from which the search starts at a high-level entry point and navigates through progressively denser graph layers, effectively narrowing down to the most similar neighbors. 
These optimizations ensure fast and scalable ANN searches, even for large datasets, which are necessary for scaling the system over large knowledge graphs such as Wikidata, where there are billions of entities and triples.

\subsubsection{Retrieving Relevant Subgraphs from the KG} 
\label{subs:subgraphretrieval} 
This module aims to retrieve relevant subgraphs, paths, or triples from large KGs that contain concise knowledge to augment the generation capability of LLMs for knowledge-intensive tasks. 
To fully leverage the rich semantic knowledge of KGs for augmenting LLMs, we retrieve relevant subgraphs from large KGs by using one of the following techniques. Among them, LLM-based subgraph retrieval, vector-based KG triple retrieval, and hybrid subgraph retrieval aim to directly identify the relevant subgraphs from the KG. Only BFS-based subgraph retrieval requires relevant entities retrieved from the previous phase (\S \ref{subs:entityretrieval}). 

\noindent \textbf{LLM-based Subgraph Retrieval.} In this method, we call the LLMs to directly retrieve the relevant subgraphs or paths that are relevant to answer the given question, without the previous step of identifying relevant entities first. 
The prompts in Appendix~\ref{app:prompts_llms_as_subgraph_retriever}, including few-shot examples with expected responses, are given to guide LLMs towards directly retrieving the relevant subgraphs from large-scale KGs. Similar to previous cases, there is a placeholder \{question\} that will be replaced with the real questions from our selected dataset in the prompt. 

Due to the fact that LLMs hallucinate and may also not know the KGs well, especially domain-specific KGs, the quality of the subgraphs retrieved by LLMs can often be poor. 
Similar to the LLM-based entity retrieval, the quality of the subgraphs retrieved by LLMs largely depends on the extent to which LLMs know the KGs. 
Since the commonsense and fact-based KGs or their open textual data forms might have already been utilized to pre-train LLMs, LLM-based subgraph retrieval works well for commonsense and fact-based KGs, while it might not work for domain-specific KGs.  

\noindent \textbf{Vector-based KG Triple Retrieval.} 
In this method, we focus on generating, storing, and retrieving embeddings for triples in large-scale knowledge graphs to facilitate vector similarity-based KG triple retrieval. First, the embeddings for KG entities and relationships are created by processing their labels and descriptions, fetched through the Wikidata API. These embeddings are generated using a pre-trained language model, e.g., RoBERTa. The embeddings are then normalized to unit vectors to enhance the efficiency of cosine similarity computations.
The embeddings, along with metadata such as entity and relationship identifiers, are stored in a vector database. KG triple retrieval is performed by computing cosine similarity between the question embedding and the KG triple embeddings, identifying the top-$k$ most relevant triples. 
    
\noindent \textbf{BFS-based Subgraph Retrieval.} After retrieving the relevant entities from large-scale KGs by leveraging either LLM-based entity retrieval or vector similarity-based entity retrieval, we explore the relevant subgraphs induced by these entities with the Breadth-First Search (BFS) algorithm. 
Given a set of retrieved entities $E=\{e_1,e_2,\dots,e_n\}$ from $\mathcal{KG}$, the retrieval strategy depends on $n$. When $n=1$, local BFS expands the neighborhood of the single seed entity up to three hops. When $n\geq 2$, we form entity pairs $ep=\{e_s,e_e\}\in\mathcal{EP}$ by combining the entities in $E$, where $e_s$ is the start entity and $e_e$ is the end entity, and retrieve the subgraphs of up to three hops connecting $e_s$ and $e_e$ (if any) via BFS. 
The three-hop search is a trade-off solution between the number of retrieved subgraphs for subgraph refinement and the time complexity and the quality of the retrieved subgraphs based on our empirical results. 
The BFS-based subgraph retrieval supports two BFS pipelines:(1) local BFS in memory, and (2) remote BFS via SPARQL queries.

The remote pipeline forms entity pairs in the same way as the local one.
For each pair, it sends a SPARQL query to Wikidata that searches for paths of up to three hops between $e_s$ and $e_e$. The queries are generated from predefined templates (see Appendix~\ref{app:sparql_templates}): the start and end entity identifiers are substituted into the template, intermediate entities and relations are left as variables, and \texttt{UNION} clauses cover both edge directions. The retrieved subgraphs are then refined with the same ranking scheme as in local BFS.

\subsubsection{Ranking-based Retrieval Refinement}
\label{subs:ranking}

Our methodology incorporates several ranking schemes to prioritize and rank the already retrieved paths from a knowledge graph. These schemes are designed to select the most relevant paths, ensuring that only the best candidates are integrated. This facilitates effective integration with LLMs by evaluating the retrieved paths based on the ranking of relevance and structural significance. 

One ranking scheme is frequency-based ranking, which prioritizes those paths that also contain several of the top-$k$ retrieved relations (\S\ref{subs:entityretrieval}: Vector-similarity-based Entity Retrieval). Recall that the top-$k$ retrieved relations are identified based on the similarity scores calculated between the question embedding and the relation embedding. 
To further enhance the relevance of retrieved paths, we employ a second ranking strategy that normalizes frequency-based scoring with respect to path length. 
This normalization ensures that longer paths, which inherently consist of more relations, do not disproportionately dominate the rankings. This strategy ensures that paths are ranked based on both the presence of relevant relations and structural conciseness.
Additionally, for \S\ref{subs:subgraphretrieval}: Vector-based KG Triple Retrieval, we directly use the cosine similarity between the question embedding and the triple embeddings to rank the retrieved triples.

Based on our ranking scheme, we select the top-1, top-2, or all retrieved paths/triples to augment the large language model. Our experimental results show that using the top-2 or top-1 retrieved paths/triples is generally more effective and efficient in terms of LLMs' retrieval-augmented generation compared to using all retrieved paths/triples. 

\subsubsection{Enhancing Prompts with Retrieved Subgraphs} 
\label{subs:prompt_enhance}

After retrieving the relevant subgraphs from external KGs, we leverage them as the semantic knowledge context to enhance the prompts of LLMs for augmenting the generation of the final responses for schema matching tasks.
To make LLMs fully understand the retrieved subgraph, there is an option to convert the retrieved subgraph to full text by introducing the verbalizer~\citep{wu2023retrieve}. 
Namely, the LLMs can serve as the verbalizer, in which case the retrieved subgraphs will be converted to text by calling the LLMs again. 
However, this option is not necessary and not the optimal solution either, since the precondition of calling LLMs to convert the subgraph to text is that the LLMs can fully understand it. 
Besides, some structured knowledge of the retrieved subgraphs will be lost during the conversion from triples to text. 

In our case, each retrieved subgraph is a semi-structured text with some special characters, such as arrows and operators, to represent multi-hop relations and separate multiple paths, which can still be understood and fully leveraged by LLMs. 
We feed the retrieved subgraphs within the prompts to LLMs directly and highlight the roles of the provided KG context with few-shot examples.
The prompts, including the few-shot examples and the retrieved subgraphs from KGs to guide and augment LLMs in making the correct decisions, are demonstrated in the Appendix~\ref{app:prompts_augmenting_generation_with_retrieved_subgraph}. 

As shown in the prompt, we provide examples with explanations to guide LLMs to fully leverage the provided KG context when making decisions and generating final answers for schema-matching questions in our given examples.
There are two placeholders, \{question\} and \{paths\}, that will be replaced with the real questions and the retrieved subgraphs from the selected dataset. 
Other placeholders [d1] and [d2] are descriptions for attribute1 and attribute2, respectively. 
It is worth mentioning that the KG context given in example 1 within the prompts provides extra knowledge about the attributes. 
Namely, the \texttt{date of death} is the \texttt{opposite of} the \texttt{date of birth}, which indicates that \texttt{date of death} and \texttt{date of birth} are not semantically equivalent even if they are syntactically similar.
This precise knowledge provides the exact relationships between two matching attribute pairs, which is helpful for LLMs to generate the correct answers for the given schema matching questions.

\subsubsection{Poorly-relevant Subgraphs Exclusion} 
\label{subs:instruction_tuning}
One challenge in augmenting LLMs with retrieved subgraphs from KG lies in the inclusion of irrelevant subgraphs that can mislead the model and result in erroneous outputs.
Although the ranking-based retrieval refinement designed in \S\ref{subs:ranking} can filter out the irrelevant subgraphs based on the vector similarity score, some poorly relevant subgraphs may still exist in the refined results.
The previous study~\citep{tian2025relevance} reveals that irrelevant context significantly hampers RAG performance in knowledge-intensive tasks, highlighting the critical importance of retrieving relevant and useful context for optimal results.  

Inspired by this, we introduce a rule-based instruction tuning strategy to further exclude poorly relevant subgraphs from the previously retrieved and ranked subgraphs and mitigate the LLM's hallucinations. As shown in the prompt, we design an instruction to prompt (detailed in Appendix~\ref{app:rule-based_it_for_poorly-relevant_subgraph_exclusion}) LLMs with examples in the manner of few-shot learning, which aims to guide LLMs to analyze the given schema matching question step by step and alleviate hallucinations by excluding the poorly relevant subgraph when making the decision. 

In particular, we instruct the LLM to assign a relevance score between 0 and 10 to each given schema matching question and the provided retrieved subgraph from the external KG, based on few-shot examples with the assigned relevance score and explanation for scoring. 
We prompt and guide LLMs to only consider the provided and retrieved knowledge graph context if the relevance score is greater than or equal to 7, thereby filtering out distracting, poorly relevant subgraphs and ensuring that only high-value triples augment the model for schema matching. 
It is worth noting that the rule-based instruction tuning for poorly relevant subgraph exclusion in \S~\ref{subs:instruction_tuning} is by default disabled in KG-RAG4SM in the main evaluation, unless otherwise stated, because it was intentionally designed for KG-RAG4SM with a domain-specific KG.
In \S\ref{sec:robust_exp}, we provide experimental results on how instruction-tuning performs for improving the truthfulness of KG-RAG4SM even in the presence of retrieved poorly relevant subgraphs.

\subsection{KG-RAG4SM} \label{sec3.3}
As introduced in \S \ref{sec3.2}, we have designed multiple knowledge retrieval methods to retrieve the relevant subgraphs from a large-sized KG to augment LLMs for schema matching. 
There are mainly four pipelines for our KG-RAG4SM with different choices of knowledge retrieval methods: \textbf{(1)} Vector-based KG triples retrieval; \textbf{(2)} Vector-based Entity Retrieval + BFS; \textbf{(3)} LLM-based Entity Retrieval + BFS; \textbf{(4)} LLM-based subgraph retrieval.

We provide details of the two pipelines that perform the best empirically for our KG-RAG4SM: \textbf{(1)} Vector-based KG triples retrieval; \textbf{(2)} Vector-based Entity Retrieval + BFS.  
The pipelines with other choices of knowledge retrieval are examined in the ablation study (\S \ref{sec:ablation}).

\noindent \textbf{Vector-based KG Triples Retrieval.} In this pipeline shown earlier in Figure \ref{fig:kgrag4sm_method}, the KG triple embeddings and the question embeddings are created by using the pre-trained language model. Then, the top-$k$ relevant KG triples are retrieved and refined by ranking based on vector similarity.

\noindent \textbf{Vector-based Entity Retrieval + BFS.} In this pipeline, vector-based entity and relation retrieval and BFS graph traversal are leveraged to retrieve the relevant knowledge from the large-sized knowledge graph $\mathcal{KG}$. Then these retrieved and refined subgraphs are employed to augment the LLMs for answering the questions $q$ of the schema matching $\theta$; the algorithm of this pipeline for our KG-RAG4SM is demonstrated below.

\begin{algorithm}
\caption{KG-RAG4SM with Vector-based Entity Retrieval and BFS-based Subgraph Retrieval}
\label{alg:kg-rag4sm}
\begin{algorithmic}[1]
\small

\State \noindent \textbf{Input:} Schema matching question $q$ of task $\theta$, entity vector index $\mathcal{I_E}$, relation vector index $\mathcal{I_R}$, knowledge graph $\mathcal{KG}$, LLM

\State \noindent \textbf{Output:} Answer $a$ for the given question $q$

\State $a \gets \emptyset$, $E \gets \emptyset$, $P, P^{\prime} \gets \emptyset$, $D_{max} \gets 3$
\Comment{Initialize parameters.}

\State $E \gets \text{vector\_similarity\_based\_entity\_retrieval}(q,\mathcal{I_E})$
\Comment{Retrieve relevant entities from the vector index.}

\If{$|E| = 0$}
    \State $P \gets \emptyset$
    \Comment{No retrieved entities.}

\ElsIf{$|E| = 1$}

    \State Let $E = \{es\}$
    \Comment{Single retrieved entity.}

    \If{$\text{supports\_single\_seed}(\mathcal{KG})$}
        \State $P \gets \text{BFS}(es,\mathcal{KG},D_{max})$
        \Comment{Expand the seed entity using local BFS.}
    \Else
        \State $P \gets \emptyset$
    \EndIf

\ElsIf{$|E| > 1$}

    \State $\mathcal{EP} \gets \text{combinations}(E,2)$
    \Comment{Construct entity pairs.}

    \For{$ep_i = \{es_i,ee_i\} \in \mathcal{EP}$}

        \State $P \gets P \cup \text{BFS} (ep_i,\mathcal{KG},D_{max})$
        \Comment{Retrieve paths of up to three hops between entity pairs.}

    \EndFor

\EndIf

\If{$|P| > 1$}

    \State $R \gets \text{vector\_similarity\_based\_relation\_retrieval}(q,\mathcal{I_R})$
    \Comment{Retrieve relevant relations from the vector index.}

    \State $P^{\prime} \gets \text{path\_refined\_with\_ranking}(P,R)$
    \Comment{Rank and refine the retrieved paths.}

\Else

    \State $P^{\prime} \gets P$

\EndIf

\State $a \gets \text{kg-rag4sm}(\text{LLM},q,P^{\prime})$
\Comment{Generate the answer using refined KG context.}

\State \Return $a$

\end{algorithmic}
\end{algorithm}

As shown in Algorithm~\ref{alg:kg-rag4sm}, we first retrieve relevant entities via vector similarity. If only one entity is retrieved, local BFS expands its neighborhood up to three hops, whereas remote SPARQL-based BFS is not applicable because it requires at least two entities. If multiple entities are retrieved, we form entity pairs and retrieve the connecting paths of up to three hops using either local BFS or SPARQL-based bounded-hop queries. The retrieved paths are then refined by relation-based ranking and included in the LLM prompt for schema matching prediction.

\section{Experimental Results} 
\label{sec4}
We empirically evaluate the performance of the proposed KG-RAG4SM method by comparing it against the state-of-the-art schema matching approaches from two domains: pre-trained language models (PLMs) and large language models (LLMs). 
We conduct experiments using three widely used schema matching benchmark datasets and a self-created dataset from the real-world healthcare database. To evaluate the cross-domain transferability of KG-RAG4SM, we additionally include two non-healthcare datasets, \textsf{Bank} and \textsf{IMSA}.
Our code and datasets are available at \textbf{Github\footnote{https://github.com/machuangtao/KG-RAG4SM}}.

We aim to investigate the following research questions.
\begin{itemize}[noitemsep, leftmargin=0.3cm, topsep=0pt]
\item \textbf{Exp-1: Effectiveness.} How does our method compare with the state-of-the-art (SOTA) models for schema matching in terms of effectiveness?
\item \textbf{Exp-2: Efficiency.} How does our method compare with the SOTA models for schema matching in terms of efficiency?
 \item \textbf{Abl-1: Subgraph Retrieval.} Which subgraph retrieval techniques work better for KG-RAG4SM in terms of both effectiveness and efficiency to generate high-quality results for schema matching?
\item \textbf{Abl-2: Backbone LLMs.}  Which LLMs perform better in KG-RAG4SM, e.g., do fewer-parameter LLMs work better than large-size LLMs for KG-RAG4SM?
\item \textbf{Ext-1: Scalability across KG Domains.} Which KGs work better in KG-RAG4SM, e.g., do domain-specific KGs work better than commonsense KGs for KG-RAG4SM?
\item  \textbf{Ext-2: Truthfulness of KG-RAG4SM.} Does the rule-based relevance scoring and instruction-tuning work for improving the truthfulness of KG-RAG4SM?
\item \textbf{Ext-3: Ablation Study on Subgraph Refinement in KG-RAG4SM and Comparison with SOTA GraphRAG.} We investigate the following two research questions: (Ext-3a)  What is the isolated contribution of the similarity ranking-based subgraph refinement mechanism to the performance of KG-RAG4SM?
(Ext-3b) How does the complete KG-RAG4SM framework perform relative to the state-of-the-art GraphRAG baseline?
\item \textbf{Case: Hallucinations.}  How does KG-RAG4SM mitigate the hallucinations of LLMs for complex schema matching in a real-world scenario? 
\end{itemize}

\subsection{Experimental Setup}
\subsubsection{Datasets}

Table~\ref{tab2} presents the statistics of the datasets used in our experiments. \textsf{MIMIC}\footnote{https://mimic.mit.edu}, \textsf{Synthea}\footnote{https://synthea.mitre.org}, and \textsf{CMS}\footnote{https://github.com/OHDSI/ETL-CMS} are extracted from the challenging \textsf{OMAP} schema-level matching benchmark~\citep{zhang2021smat}. \textsf{OMAP} maps three
different healthcare data models (\textsf{MIMIC}~\citep{mimic}, \textsf{Synthea}~\citep{synthea}, and \textsf{CMS}~\citep{cms}) with the \textsf{OMOP CDM} (Observational Medical Outcomes Partnership Common Data Model) standard~\citep{ohdsi2019book}. \textsf{EMED} is our newly created schema matching benchmark by mapping a dataset from the \textsf{e-MedSolution} domain to the \textsf{OMOP CDM} model. 
The \textsf{EMED} dataset was created and annotated with the following pipeline: \textbf{(1)} Data Preprocessing. The textual description of each table and attribute name originates from the database design document of the e-MedSolution system, which was then translated from Hungarian to English. \textbf{(2)} Matching Pair Generation. The matching pairs were generated based on the Cartesian product of the table and attribute names in the e-MedSolution data model and the subset of the tables and attributes from the OMOP CDM data model. \textbf{(3)} Data Annotation. The matching pair was initially annotated manually based on the translated textual descriptions of the table and attribute names in the source and target data models and then partially reviewed and verified by a Hungarian native speaker for consistency.
The datasets of \textsf{Bank} were derived from LLMATCH~\citep{WangLLMATCH25}.
A brief introduction to the data models from the source and target schemata is given below. 
\begin{itemize}[noitemsep, leftmargin=0.3cm, topsep=0pt]
    \item \textsf{MIMIC} is a publicly available 
    ICU relational database from the Beth Israel Deaconess Medical Center. 
    \item \textsf{Synthea} is an open-source dataset that captures the medical history of $\approx$ 1 million Massachusetts synthetic patients. 
    \item \textsf{CMS} is a set of realistic medicare claims data generated from medicare claims synthetic public use files (SynPUFs).
    \item \textsf{EMED} is a dataset constructed by us based on the \textsf{e-MedSolution}, which is a healthcare information system (HIS) that has been applied in various healthcare institutions and medical universities in Hungary. 
     \item \textsf{OMOP CDM} is a standardized and common healthcare data model for analysis of disparate observational databases that provides a comprehensive view of clinical and healthcare data for patients and medical staff.
     \item \textsf{Bank} is a financial schema matching dataset that includes matching pairs between data models of two banks. The pairs ({\em bank1}-{\em bank2}) were anonymized and approved for release with industry partners.
     \item \textsf{IMSA} is an entertainment data model that includes movie matching pairs between IMDB and SAKILA, which were manually annotated by experts.
\end{itemize}

\begin{table}
\caption{Statistics of Dataset for Schema Matching.}
\label{tab2}
\small
\begin{center}
\begin{tabular}{c|c|c|rr|rr|r|r}
\toprule
\multirow{2}{*}{Dataset}  & \multirow{2}{*}{Mapping}         & \multirow{2}{*}{Domain} & \multicolumn{2}{c|}{Source}                  & \multicolumn{2}{c|}{Target}                  & \multirow{2}{*}{\makecell{\#Matching \\ Pairs}} & \multirow{2}{*}{\makecell{\#Matched \\ Pairs}} \\ \cline{4-7}
                          &                                  &                         & \multicolumn{1}{r|}{\#Tab.} & \#Attr. & \multicolumn{1}{r|}{\#Tab.} & \#Attr. &                                   &                                  \\ \midrule
\textsf{MIMIC}              & MIMIC-OMOP                       & healthcare              & \multicolumn{1}{r|}{25}      & 240          & \multicolumn{1}{r|}{21}       & 267& 64,080                            & 129                              \\ \hline
\textsf{Synthea}            & Synthea-OMOP                     & healthcare              & \multicolumn{1}{r|}{12}       & 111          & \multicolumn{1}{r|}{21}       & 267& 29,637                            & 105                              \\ \hline
\textsf{CMS}                & CMS-OMOP                         & healthcare              & \multicolumn{1}{r|}{5}        & 96           & \multicolumn{1}{r|}{21}       & 267& 25,632                            & 196                              \\ \hline
\textsf{EMED}               & EMED-OMOP                        & healthcare              & \multicolumn{1}{r|}{55}       & 804& \multicolumn{1}{r|}{7}       & 101& 81,204                            & 239                              \\ \hline
\textsf{Bank} & \multicolumn{1}{r|}{Bank1-Bank2} & finance                 & \multicolumn{1}{r|}{9}        & 27           & \multicolumn{1}{r|}{9}        & 36           & 972          & 17          \\ \hline
\textsf{IMSA} & \multicolumn{1}{r|}{IMDB-SAKILA} & movie                   & \multicolumn{1}{r|}{7}        & 39           & \multicolumn{1}{r|}{16}      & 90& 2,379        & 50           \\ 
\bottomrule
\end{tabular}  
\end{center}
\end{table}

The number of tables (\#Tab.) and attributes (\#Attr.) from the source data models differs from those in the target data model due to structural heterogeneity. 
The number of matching pairs for each dataset is the combination of the attributes in each source data model and the attributes in the target 
data model. 
Some unlabeled pairs in the \textsf{IMSA} dataset are further filtered out, yielding fewer matching pairs than the original pair counts.
The \#matched pairs count the number of positive matchings between the source data model and the target data model.

It is worth mentioning that the traditional supervised SOTA methods, such as Unicorn~\citep{tu2023unicorn} and SMAT~\citep{zhang2021smat}, which learn schema matching following a supervised learning paradigm, split each dataset into training, validation, and testing sets based on the split ratio of 0.70, 0.15, and 0.15. 
In contrast, our KG-RAG4SM performs schema matching based on the provided examples and prompts via few-shot learning and without any retraining of LLMs.
To ensure a fair comparison, we evaluate our KG-RAG4SM on the test sets of \textsf{MIMIC}, \textsf{Synthea}, and \textsf{CMS} directly from SMAT~\citep{zhang2021smat}, and the test sets of \textsf{EMED}, \textsf{Bank}, and \textsf{IMSA} with the same data split ratio as the above methods.

\subsubsection{Knowledge Graph}

The Wikidata\footnote{https://www.wikidata.org} knowledge graph is primarily selected as the external knowledge graph to enhance schema matching, since Wikidata contains general and common knowledge in the form of structured data to support Wikipedia. 
The UMLS KG is also selected to demonstrate the scalability across KG domains for the proposed KG-RAG4SM.
Because the Wikidata KG is updated dynamically, we report the statistics of Wikidata KG directly from~\citep{haller2022an} and the statistics of Wikidata5M~\citep{WangGZZLLT21} in Table~\ref{tab3}.
\begin{itemize}[noitemsep, leftmargin=0.3cm, topsep=0pt]
    \item Wikidata dumps\footnote{https://dumps.wikimedia.org} is a full-size Wikidata KGs containing all Wikidata entities in a single JSON or RDF dump. The Wikidata dumps can be accessed using the Wikidata API\footnote{https://www.wikidata.org/w/api.php}, which allows access to specific data from the Wikidata dump by making HTTP requests. 
    \item Wikidata5M\footnote{https://deepgraphlearning.github.io/project/wikidata5m} is the core subgraph of Wikidata, containing high-frequency entities from Wikidata with their corresponding relations and triples. 
    \item  UMLS\footnote{https://uts.nlm.nih.gov/uts} is a healthcare and biomedical metathesaurus that contains over 3.8 million medical concepts and 78 million relations in the form of triples.
\end{itemize}

We use the Wikidata API\footnote{https://www.wikidata.org/w/api.php} to verify identifiers and retrieve missing entity and relation labels, and the SPARQL endpoint\footnote{https://query.wikidata.org/sparql} for remote BFS retrieval. Due to limited computing resources, we use Wikidata5M to generate the embeddings for vector-based entity, relation, and triple retrieval, and as the graph for local in-memory BFS. With sufficient resources, our approach extends directly to the full Wikidata KG by loading its official dumps.

\begin{table}
\caption{Statistics of Wikidata KG}\label{tab3}
\small
\begin{center}
\begin{tabular}{c|c|c|c} \toprule
KGs & \#Entity & \#Relation & \#Triple\\ \midrule
Wikidata & 4,232,253,847 & 38,867 & 1,693,668,039 \\ \hline
Wikidata5M & 4,594,485 & 822 & 20,624,575 \\ \bottomrule
\end{tabular}
\end{center}
\end{table}

\subsubsection{SOTA Models}
We compare the performance of KG-RAG4SM against two categories of SOTA schema matching models: (1) \textbf{PLMs}, such as SMAT~\citep{zhang2021smat} and Unicorn~\citep{tu2023unicorn}; (2) \textbf{LLMs}, for example, Jellyfish~\citep{zhang2024jellyfish}, and the details are given below.
\begin{itemize}[noitemsep, leftmargin=0.3cm, topsep=0pt]
    \item \textbf{SMAT} employs an attention over attention (AOA)~\citep{cui2017attention} mechanism and BiLSTM networks to model the correlation between the attribute name and its description to obtain a better sentence representation and subsequently improve schema matching.
    \item \textbf{Unicorn} is a unified multi-tasking model for supporting several matching tasks in data integration, in which the pre-trained language models and the Mixture-of-Experts (MoE) modules are introduced to encode and map various matching tasks for knowledge sharing. 
   \item \textbf{Jellyfish} adopts an instruction-tuned LLM for data preprocessing that enables optional knowledge injection during inference by instruction tuning with reasoning based on the domain knowledge.  
\end{itemize}
Although there are some other recent methods, such as ReMatch~\citep{sheetrit2024rematch} and Matchmaker~\citep{seedat25bootstrapping}, that leverage traditional RAG over LLMs to address schema matching, they only augment LLMs by retrieving the relevant knowledge from the unstructured textual documents that reside within the schema.
KcMF~\citep{xu2024kcmf} retrieves the knowledge from the self-built knowledge set, while our method retrieves the knowledge from large-size commonsense and domain-specific KGs. 
These methods are not selected for comparison due to inaccessible source code and the difference in retrieval augmentation.  
Our methods differ from the above methods; we retrieve the relevant subgraphs from the large-size commonsense, fact-based, and domain-specific KGs in a real-world scenario, which is external knowledge compared to the textual documents constructed based on the descriptions of the database table and attribute. 

\subsubsection{Evaluation Metrics}

Considering that schema matching is treated as a classification task in our case, 
\texttt{Precision} \texttt{(P)}, \texttt{Recall(R)}, and \texttt{F1} score are selected to evaluate the performance of the proposed and competing methods. 
Precision measures the proportion of true positive (\texttt{TP}) matches over all returned matches (i.e., \texttt{True Positive (TP)} + \texttt{False Positive (FP)}), while \texttt{recall(R)} measures the proportion of true positive matches over all positive matches  (i.e., \texttt{True Positive (TP)} + \texttt{False Negative (FN)}). \texttt{F1} score is a balance between precision and recall.

$$\texttt{P}= \frac{\texttt{TP}}{\texttt{TP}+\texttt{FP}} \quad \texttt{R}= \frac{\texttt{TP}}{\texttt{TP}+\texttt{FN}} \quad \texttt{F1}=2\cdot \frac{\texttt{P} \cdot \texttt{R}}{\texttt{P}+\texttt{R}}$$

We also evaluate the efficiency of the subgraph retrieval and answer generation. 
The retrieval time measures the time of various subgraph retrieval methods, for instance, vector-based entity retrieval, BFS, subgraph ranking and refinement, while the generation time measures the generation time of LLMs in making their decisions and generating the final answer for the given schema matching questions.

\subsubsection{Implementation}
The details of the KG-RAG4SM implementations and the hyperparameters are given as follows.

\textbf{System.} We run the models on a remote server with Ubuntu 22.04, 60$\times$ Intel(R) Xeon (Ice Lake) CPUs @2.8GHz, 2$\times$ NVIDIA A10 GPUs with CUDA, 100GB RAM, and 1024GB HDD.
We install and manage PyTorch and its dependency packages using Anaconda\footnote{https://anaconda.org/anaconda}.

We deployed a separate virtual machine configured with Ubuntu 22.04.5, 32$\times$ Intel Xeon Skylake vCPUs, and 64~GB RAM to implement the local in-memory BFS graph traversal. This environment was used specifically to profile the memory and storage footprint, adjacency-cache construction cost, graph-loading and initialization time, and hardware requirements for maintaining Wikidata5M locally.
Detailed deployment measurements of the local in-memory BFS implementation are reported in Appendix~\ref{sec:graph-retrieval-deployment-efficiency}.

\textbf{Backbone LLMs}. The GPT-4o-mini from OpenAI is mainly selected as the backbone LLM in our effectiveness and efficiency experiments.
We utilize the GPT model using the official API\footnote{https://platform.openai.com/} from OpenAI and load the Mistral\footnote{https://huggingface.co/mistralai} and Jellyfish\footnote{https://huggingface.co/NECOUDBFM} models from HuggingFace. 

\textbf{Vector Representations.} The sentence transformers framework with RoBERTa-base\footnote{https://huggingface.co/FacebookAI/roberta-base} is leveraged to derive the vector representations of the given schema-matching questions.
Similarly, entities, relations, and triples from the selected knowledge graphs are used to obtain their vector representations with the same embedding model and vector dimensions and then organized with the ChromaDB embedding database\footnote{https://github.com/chroma-core/chroma}. 

\textbf {Vector-similarity-based Retrieval.} The cosine distance between each question vector and entity, relation, and triple vector from the KG is calculated and ranked with respect to the given schema-matching question to retrieve the relevant entities, relations, and triples from the KG. ChromaDB is employed to implement efficient vector search since it leverages optimized vector indexing techniques to speed up similarity searches significantly.

\textbf{Ranking-based Subgraph Refinement.}  
The top-$1$ and top-$2$ most relevant subgraphs are refined by
our ranking scheme (\S\ref{subs:ranking}).

\textbf{Hyperparameters.} To make a fair evaluation, the common hyperparameters across different models are set to the same values as shown in Table~\ref{tab4}.

\begin{table}
\caption{The Value of the Common Hyperparameters.} \label{tab4}
\small
\begin{center}
\begin{tabular}{c|p{10cm}|c} \toprule
Parameter Name & Description & Value  \\ \midrule
max\_new\_tokens & The maximum number of generated output tokens. & 4096  \\ \hline
temperature & The entropy size of LLMs to control the trade-off between predictability and creativity of the generation and response. & 0.5  \\ \hline
top-$k$ & It is a parameter that enforces deterministic greedy decoding by selecting the k highest-probability tokens at each step. & 1  \\ \hline
top-$p$ & It is a parameter that selects the smallest set of tokens whose cumulative probability meets or exceeds the probability $p$. & 0.9  \\ \hline
embed\_dim & The dimension of the text vector space. & 768 \\ \hline
batch\_size & The size of the batch data for each iteration. & 32 \\ \bottomrule 
\end{tabular}
\end{center}
\end{table}

\subsection{Effectiveness and Efficiency Results}
This section reports the experimental results, highlighting the best scores in bold and the second-best scores underlined. 
These results are further analyzed and discussed to address the aforementioned research questions by comparing our KG-RAG4SM against baselines on the datasets. 
\subsubsection{Exp-1: Effectiveness} 
\textbf{\textit{How does our method compare with the state-of-the-art (SOTA) models for schema matching in terms of effectiveness?}}

\noindent \underline{\textit{Evaluating KG-RAG4SM.}} 
We implement the KG-RAG4SM with the following settings: \textbf{(1) Backbone LLM}: GPT-4o-mini and Jellyfish-8B are selected as the backbone LLMs. \textbf{(2) Subgraph Retrieval:} Subgraphs retrieved by vector-based KG triples retrieval with the refined top-2 paths. 
We keep the same hyperparameters (Table~\ref{tab4}) and prompts in KG-RAG4SM and LLM-based schema matching with Jellyfish-8B and GPT-4o-mini. 

\noindent \underline{\textit{Evaluating other Models.}}
We have reproduced the results of SOTA models based on their implementation with the same common hyperparameters. 
We report our reproduced results of SMAT and Unicorn, in which Glove and DeBERTa~\citep{he2021deberta} are selected as the encoders for SMAT and Unicorn, respectively.
We train SMAT on the training set of each dataset individually, whereas we train Unicorn jointly across the training sets of all selected datasets, keeping the same max. input text length and max. number of epochs with their original hyperparameters, respectively.
Namely, the max. input text length and the max. number of epochs for Unicorn are 128 and 10, while those for SMAT are 240 and 10, respectively.
For Unicorn, we also fine-tune the pre-trained UnicornPlus\footnote{https://huggingface.co/RUC-DataLab/unicorn-plus-v1} model on the training set of each selected dataset separately. 
For Jellyfish, we load their fine-tuned LLMs from  HuggingFace\footnote{https://huggingface.co/NECOUDBFM/Jellyfish-8B} to generate the results of schema matching on our datasets.     

\noindent \underline{\textit{Results Analysis.}} 
We first evaluate and report the effectiveness of our KG-RAG4SM model compared to the selected LLMs, i.e., GPT-4o-mini and Jellyfish-8B, across datasets. 
The overall comparison results are reported in Table~\ref{tab5}, where the best ones are highlighted with bold values.

\begin{table}
\caption{Effectiveness of KG-RAG4SM with the top-2 ranked subgraphs by using vector-based KG triples retrieval in comparison to LLM models.} \label{tab5}
\small
\begin{center}
\begin{tabular}{c|c|cc||cc}
\toprule
\multirow{2}{*}{Dataset}                       & \multirow{2}{*}{\begin{tabular}[c]{@{}c@{}}Metric\\ (\%)\end{tabular}} & \multicolumn{2}{c||}{KG-RAG4SM}                                                                                                                                                   & \multicolumn{2}{c}{LLM}                                                                                                                                               \\ \cline{3-6} 
                                               &                                                                        & \multicolumn{1}{c|}{GPT-4o-mini} & Jellyfish-8B & \multicolumn{1}{c|}{GPT-4o-mini} & Jellyfish-8B \\ \midrule
\multirow{3}{*}{\textsf{MIMIC}}   & \texttt{P}                                                & \multicolumn{1}{c|}{6.25}                                                                       & \multicolumn{1}{c||}{\textbf{9.52} ($\uparrow$39.79\%)}                                                                            & \multicolumn{1}{c|}{6.25}                                                             & 6.81                                                                          \\ \cline{2-6} 
                                               & \texttt{R}                                                & \multicolumn{1}{c|}{33.33}                                                                      &  \multicolumn{1}{c||}{66.66($\downarrow$33.34\%)}                                                                           & \multicolumn{1}{c|}{33.33}                                                            & \textbf{100 }                                                                          \\ \cline{2-6} 
                                               & \texttt{F1}                                               & \multicolumn{1}{c|}{10.52}                                                                      &  \multicolumn{1}{c||}{\textbf{16.66}($\uparrow$30.56\%)}                                                                           & \multicolumn{1}{c|}{10.52}                                                            & 12.76                                                                         \\ \hline \hline
\multirow{3}{*}{\textsf{Synthea}} & \texttt{P}                                                & \multicolumn{1}{c|}{20.00($\uparrow$33.33\%)}                                                                      &  \multicolumn{1}{c||}{\textbf{26.31}($\uparrow$150.10\%)}                                                                           & \multicolumn{1}{c|}{15.00}                                                            & 10.52                                                                         \\ \cline{2-6} 
                                               & \texttt{R}                                                & \multicolumn{1}{c|}{36.36($\uparrow$33.33\%)}                                                                     &       \multicolumn{1}{c||}{\textbf{45.45}($\uparrow$25.00\%)}                                                                       & \multicolumn{1}{c|}{27.27}                                                            & 36.36                                                                         \\ \cline{2-6} 
                                               & \texttt{F1}                                               & \multicolumn{1}{c|}{25.80($\uparrow$33.33\%)}                                                                     &    \multicolumn{1}{c||}{\textbf{33.33}($\uparrow$104.23\%)}                                                                          & \multicolumn{1}{c|}{19.35}                                                            & 16.32                                                                         \\ \hline \hline
\multirow{3}{*}{\textsf{CMS}}     & \texttt{P}                                                & \multicolumn{1}{c|}{\textbf{52.38}($\uparrow$17.87\%)}                                                                      &    \multicolumn{1}{c||}{23.33($\downarrow$22.23\%)}                                                                         & \multicolumn{1}{c|}{44.44}                                                            & 30.00                                                                         \\ \cline{2-6} 
                                               & \texttt{R}                                                & \multicolumn{1}{c|}{44.00($\downarrow$8.33\%)}                                                                      &      \multicolumn{1}{c||}{28.00($\downarrow$22.22\%)}                                                                       & \multicolumn{1}{c|}{\textbf{48.00}}                                                            & 36.00                                                                         \\ \cline{2-6} 
                                               & \texttt{F1}                                               & \multicolumn{1}{c|}{\textbf{47.82}($\uparrow$3.62\%)}                                                                      &   \multicolumn{1}{c||}{25.45($\downarrow$22.22\%)}                                                                          & \multicolumn{1}{c|}{46.15}                                                            & 32.72                                                                         \\ \hline \hline
\multirow{3}{*}{\textsf{EMED}}    & \texttt{P}                                                & \multicolumn{1}{c|}{\textbf{3.03}($\uparrow$1.68\%)}                                                              &      \multicolumn{1}{c||}{0}                                                                       & \multicolumn{1}{c|}{2.98}                                                             & 0                                                                             \\ \cline{2-6} 
                                               & \texttt{R}                                                & \multicolumn{1}{c|}{\textbf{50.00}}                                                                      &    \multicolumn{1}{c||}{0}                                                                         & \multicolumn{1}{c|}{\textbf{50.00}}                                                            & \multicolumn{1}{c}{0}                                                          \\ \cline{2-6} 
                                               & \texttt{F1}                                               & \multicolumn{1}{c|}{\textbf{5.71}($\uparrow$1.42\%)}                                                              &      \multicolumn{1}{c||}{0}                                                                       & \multicolumn{1}{c|}{5.63}                                                             & \multicolumn{1}{c}{0}                                                          \\ \hline \hline
\multirow{3}{*}{\textsf{Bank}}    & \texttt{P}                                                & \multicolumn{1}{c|}{\textbf{33.33}($\uparrow$50.00\%)}                                                              &      \multicolumn{1}{c||}{28.57($\uparrow$28.58\%)}                                                                       & \multicolumn{1}{c|}{22.22}                                                             & 22.22\\ \cline{2-6} 
                                               & \texttt{R}                                                & \multicolumn{1}{c|}{100}                                                                      &    \multicolumn{1}{c||}{100}                                                                         & \multicolumn{1}{c|}{\textbf{}100}                                                            & \multicolumn{1}{c}{100}                                                          \\ \cline{2-6} 
                                               & \texttt{F1}                                               & \multicolumn{1}{c|}{\textbf{50.00}($\uparrow$37.51\%)}                                                              &      \multicolumn{1}{c||}{44.44($\uparrow$22.22\%)}                                                                       & \multicolumn{1}{c|}{36.36}                                                             & \multicolumn{1}{c}{36.36}                                                          \\  \hline \hline
\multirow{3}{*}{\textsf{IMSA}}    & \texttt{P}                                                & \multicolumn{1}{c|}{\textbf{100}($\uparrow$200.03\%)}                                                              &      \multicolumn{1}{c||}{\textbf{100}($\uparrow$300\%)}                                                                       & \multicolumn{1}{c|}{33.33}                                                             & 25.00\\ \cline{2-6} 
                                               & \texttt{R}                                                & \multicolumn{1}{c|}{16.67}                                                                      &    \multicolumn{1}{c||}{16.67}                                                                         & \multicolumn{1}{c|}{16.67}                                                            & \multicolumn{1}{c}{16.67}                                                          \\ \cline{2-6} 
                                               & \texttt{F1}                                               & \multicolumn{1}{c|}{\textbf{28.57}($\uparrow$28.58\%)}                                                              &      \multicolumn{1}{c||}{\textbf{28.57}($\uparrow$42.85\%)}                                                                       & \multicolumn{1}{c|}{22.22}                                                             & \multicolumn{1}{c}{20.00}                                                          \\ 
                                               
\bottomrule
\end{tabular}    
\end{center}
\end{table}

Our KG-RAG4SM method overall outperforms the selected LLM competing models in terms of both \texttt{P} and \texttt{F1} score across the datasets, except for GPT-4o-mini as a backbone model on the \textsf{MIMIC} dataset and Jellyfish-8B on the \textsf{CMS} dataset.
Notably, KG-RAG4SM contributes \textbf{39.79\%} and \textbf{30.56\%} improvement in \texttt{P} and \texttt{F1} score on \textsf{MIMIC}, \textbf{150.10\%} and \textbf{104.23\%} improvement on \textsf{Synthea}, \textbf{28.58\%} and \textbf{22.22\%} improvement on \textsf{Bank}, and \textbf{300.00\%} and \textbf{42.85\%} improvement on \textsf{IMSA} with Jellyfish-8B as the backbone LLM.
Our KG-RAG4SM also achieves \textbf{17.87\%} and \textbf{3.62\%} improvement of \texttt{P}, \texttt{F1} score on \textsf{CMS}, \textbf{1.68\%} and \textbf{1.42\%} improvement on \textsf{EMED}, \textbf{50.00\%} and \textbf{37.51\%} improvement on \textsf{Bank}, and \textbf{200.03\%} and \textbf{28.58\%} improvement on \textsf{IMSA} with GPT-4o-mini as the backbone LLM.

Even if the KG-RAG4SM does not yield improvement in \texttt{R} on the \textsf{MIMIC} and \textsf{CMS} datasets, we notice that 66.66\% \texttt{R} could be better than 100\% \texttt{R} in practice, since 100\% \texttt{R} seems overfitted with no false negatives and relatively higher false positives, indicating that the baseline Jellyfish LLM is more prone to say 'YES' than 'NO'  irrespective of specific schema matching questions. 
We also notice a performance decrease of KG-RAG4SM with Jellyfish-8B as the backbone LLM on the \textsf{CMS} dataset. The reason might be the catastrophic forgetting of fine-tuned LLMs~\citep{luo2023empirical} and the knowledge conflicts when integrating the new factual knowledge with the fine-tuned LLMs on the unseen \textsf{CMS} dataset. 
This aligns with previous findings~\citep{barnett2024fine}, indicating that RAG with a fine-tuned model negatively impacted the quality of the generated answers.
It is worth mentioning that Jellyfish reports 0 across the metrics on \textsf{EMED}, the reasons are twofold: \textbf{(1)} Jellyfish is a fine-tuned LLM model, especially for data preprocessing tasks, which is easy to underfit for the specific schema matching tasks on an unknown dataset; \textbf{(2)} the imbalance of the ground-truth answers for schema matching in \textsf{EMED} dataset.
\textit{The overall results suggest that KG-RAG4SM is a more effective method augmented with retrieved knowledge from the large KG than the baseline LLMs without knowledge augmentation for schema matching. } 

Next, we evaluate and report the effectiveness of our KG-RAG4SM compared to the selected PLMs, i.e., Unicorn and SMAT. 
The overall comparison results are reported in Table~\ref{tab6}, where the best scores and second-best scores are highlighted in bold and underlined.
\begin{table}
\caption{Effectiveness of KG-RAG4SM with the top-2 ranked subgraphs by using vector-based KG triples retrieval in comparison to PLM models.} \label{tab6}
\small
\begin{center}
\begin{tabular}{c|c|c||c|c||c}
\toprule
\multirow{2}{*}{Dataset}                       & \multirow{2}{*}{Metric (\%)} & \multirow{2}{*}{KG-RAG4SM} & \multicolumn{2}{c||}{Unicorn}                       & SMAT    \\ \cline{4-6} 
                                               &                                                                        &                                                                                    & \multicolumn{1}{c|}{Trained} & Fine-tuned & Trained \\ \midrule
\multirow{3}{*}{\textsf{MIMIC}}   & \texttt{P}                                                & \textbf{6.25}                                                                      & \multicolumn{1}{c|}{\underline{0.04}}                 & 0                & \underline{0.04}\\ \cline{2-6} 
                                               & \texttt{R}                                                & 33.33& \multicolumn{1}{c|}{\underline{99.99}}                 & 0               & \textbf{100}\\ \cline{2-6} 
                                               & \texttt{F1}                                               & \textbf{10.52}                                                                     & \multicolumn{1}{c|}{\underline{0.09}}                 & 0                & \underline{0.09}\\ \hline \hline
\multirow{3}{*}{\textsf{Synthea}} & \texttt{P}                                                & \textbf{20.00}                                                                     & \multicolumn{1}{c|}{0.37}                 & 0                & \underline{11.82}               \\ \cline{2-6} 
                                               & \texttt{R}                                                & 36.36& \multicolumn{1}{c|}{\underline{99.99}}                 & 0               & \textbf{100}             \\ \cline{2-6} 
                                               & \texttt{F1}                                               & \textbf{25.80}                                                                     & \multicolumn{1}{c|}{0.74}                 & 0               &  \underline{21.15}            \\ \hline \hline
\multirow{3}{*}{\textsf{CMS}}     & \texttt{P}                                                & \underline{52.38}                                                                     & \multicolumn{1}{c|}{0.97}                 & \textbf{59.99}                & 31.57  \\ \cline{2-6} 
                                               & \texttt{R}                                                & 44.00                                                                              & \multicolumn{1}{c|}{\textbf{99.99}}                 & 35.99      & \underline{48.00}   \\ \cline{2-6} 
                                               & \texttt{F1}                                               & \textbf{47.82}                                                                     & \multicolumn{1}{c|}{1.93}                 & \underline{44.99}                & 38.09              \\ \hline \hline
\multirow{3}{*}{\textsf{EMED}}    & \texttt{P}                                                & \textbf{3.03}                                                                      & \multicolumn{1}{c|}{0.04}                 & 0.98                & \underline{1.25}             \\ \cline{2-6} 
                                               & \texttt{R}                                                & 50.00& \multicolumn{1}{c|}{\underline{99.99}}                 & 24.99               & \textbf{100}              \\ \cline{2-6} 
                                               & \texttt{F1}                                               & \textbf{5.71}                                                                      & \multicolumn{1}{c|}{0.09}                 & 1.88                & \underline{2.46}             \\ \hline \hline
 \multirow{3}{*}{\textsf{Bank}}& \texttt{P}                                                & \underline{33.33}& \multicolumn{1}{c|}{1.37} & \textbf{39.99}&1.36\\
 \cline{2-6} & \texttt{R}                                                & \textbf{100}& \multicolumn{1}{c|}{\underline{99.99}} & \underline{99.99}&\textbf{100}\\
                                                \cline{2-6} &  \texttt{F1} & \underline{50.00}& \multicolumn{1}{c|}{2.70} & \textbf{57.14}&2.70\\ \hline \hline
 \multirow{3}{*}{\textsf{IMSA}}& \texttt{P}       & \textbf{100}& \multicolumn{1}{c|}{1.70} & \underline{99.99}&1.70\\ 
 \cline{2-6} & \texttt{R}                                                &  16.67& \multicolumn{1}{c|}{\underline{99.99}}  & 33.33&\textbf{100}\\
 \cline{2-6} & \texttt{F1} & \underline{28.57}& \multicolumn{1}{c|}{3.32}  & \textbf{49.99}&3.32\\
 \bottomrule
\end{tabular}  
\end{center}
\end{table}

Our method outperforms trained Unicorn and SMAT in terms of \texttt{F1} score across datasets. 
A similar trend is observed against the fine-tuned Unicorn model, except in the \textsf{Bank} and \textsf{IMSA} datasets.
This can be attributed to the fact that Unicorn is pre-trained on multiple cross-domain datasets, which include entertainment-specific data highly similar to the domain of the \textsf{Bank} and \textsf{IMSA} datasets.
We can also observe that fine-tuned Unicorn models demonstrate poor performance on unseen datasets, e.g., \textsf{MIMIC}, \texttt{Synthea}, \texttt{CMS}, etc., where the domain of these datasets for pre-training is different from the fine-tuning dataset.
This reveals the inherent limitations of the fine-tuned model in terms of domain scalability and robustness, as it struggles to adapt to unseen datasets outside the domain of its training datasets.
For instance, KG-RAG4SM with GPT-4o-mini outperforms the pre-trained SMAT by \textbf{69.20\%} and \textbf{21.99\%} in terms of \texttt{P} and \texttt{F1} score on the \textsf{Synthea} dataset, respectively.
Notably, even though our results are not competitive with the pre-trained Unicorn and SMAT models in terms of \texttt{R}, the model with 100\% and 99.99\% tends to be overfitted.
As we mentioned before, the \texttt{P} and \texttt{F1} score are two crucial metrics for our schema matching case.
Additionally, the results of the fine-tuned Unicorn model in \textsf{MIMIC} and \textsf{Synthea} are 0 since there are no true positive results obtained by this model, which further verifies the previous conclusion: these models easily tend to be overfitted. 
This suggests that our method achieves the best results in terms of \texttt{P} and \texttt{F1} score across datasets. 
It is also worth mentioning that these PLM-based models need supervised learning, in which labeled data and powerful computational resources are required for training.  
Our KG-RAG4SM differs from PLM-based methods since labeled data and supervised training processes are not required in our method.
Overall, the results from Table~\ref{tab5} and Table~\ref{tab6} demonstrate the effectiveness of our method compared to the SOTA LLMs and PLM models.

\subsubsection{Exp-2: Efficiency} 
\textbf{\textit{How does our method compare with the SOTA models for schema matching in terms of efficiency?}}

\noindent \underline{\textit{Evaluation Setting.}} We measure the average question embedding time (\texttt{QET}), retrieval time(\texttt{RT}), and generation time (\texttt{GT}) per question for our KG-RAG4SM, as well as the average generation time per question for LLM-based methods.
We also present the total fine-tuning time (\texttt{FTT}) and training time (\texttt{TT}) for PLM-based methods, i.e., Unicorn and SMAT. For Unicorn, the fine-tuning time is the time of fine-tuning the model on each of the training sets separately, and the training time of each dataset is the average training time per number of matching pair in the corresponding training set of each dataset.
It is worth mentioning that the generation time of PLM-based methods is much faster; for instance, the final answers of SMAT are generated within 3 seconds on the \textsf{EMED} dataset. 
This is due to the fact that the number of parameters of PLM-based models is lower than LLMs.

\noindent \underline{\textit{Results Analysis.}} The efficiency comparison results are given in Table~\ref{tab7}, from which we observe that our method is more efficient than pre-trained SMAT and pre-trained and fine-tuned Unicorn, even if our method introduces knowledge retrieval.
Interestingly, the generation time of our method is slightly lower than the generation time of the same LLM model without RAG. The reason might be that the retrieved relevant subgraphs within KG-RAG4SM can help LLMs to locate the required knowledge easily, rather than searching the knowledge in its entire parameter space, and this further reduces the number of inferences for LLMs to generate a confident response.
This also indicates that our method with the retrieved subgraph augmentation would not induce extra time cost for generation. 
The results indicate that the main time cost of our KG-RAG4SM method is due to vector retrieval, while the time cost of question embedding is very low. In contrast to PLM-based models, our KG-RAG4SM method performs offline embeddings of KG entities, relations, and triples, which are independent of the schema matching task. 
Therefore, our approach avoids tedious supervised training and task-specific fine-tuning on labeled data. Specifically, generating these offline KG embeddings takes 20 hours for Wikidata5M entities and 26 hours for Wikidata5M triples under our system setup, which is reported separately here as a one-time cost. Overall, KG-RAG4SM demonstrates better efficiency during practical deployment than PLM baselines.

\begin{table}[bt!]
\caption{Efficiency of KG-RAG4SM with the top-2 ranked subgraphs using vector-based KG triples retrieval, compared to SOTA models.} \label{tab7}
\small
\begin{center}
\resizebox{\linewidth}{!}{
\begin{tabular}{c|ccc||cc||rrr}
\toprule
\multirow{3}{*}{Dataset} & \multicolumn{3}{c||}{KG-RAG}                                                                                                                                                                                                                                & \multicolumn{2}{c||}{LLMs}                                                                                                                                & \multicolumn{3}{c}{PLMs}                                                                                                                                                                                                                       \\ \cline{2-9} 
                         & \multicolumn{3}{c||}{KG-RAG4SM (GPT-4o-mini)}                                                                                                                                                                                                      & \multicolumn{1}{c|}{Jellyfish-8B}                                            & GPT-4o-mini   & \multicolumn{2}{c|}{Unicorn}                                                                                                                                        & SMAT                                                  \\ \cline{2-9} 
                         & \multicolumn{1}{c|}{\texttt{QET}  (Sec)} & \multicolumn{1}{c|}{\texttt{RT} (Sec)} & \texttt{GT} (Sec) & \multicolumn{1}{c|}{\texttt{GT} (Sec)} & \texttt{GT} (Sec) & \multicolumn{1}{c|}{\texttt{FTT} (Sec)} & \multicolumn{1}{c|}{\texttt{TT} (Sec)} & \texttt{TT} (Sec) \\ \midrule
\textsf{MIMIC}           & \multicolumn{1}{c|}{0.01}                                                                        & \multicolumn{1}{c|}{2.60}                                                            & 0.64                                                             & \multicolumn{1}{c|}{0.26}                                                             & 0.77                                                             & \multicolumn{1}{r|}{3389.90}                                                           & \multicolumn{1}{r|}{1627.45}                                                        & 3064.62                                                        \\ \hline
\textsf{Synthea}         & \multicolumn{1}{c|}{0.04}                                                                        & \multicolumn{1}{c|}{2.59}                                                            & 0.51                                                             & \multicolumn{1}{c|}{0.26}                                                             & 0.60                                                             & \multicolumn{1}{r|}{2039.59}                                                           & \multicolumn{1}{r|}{752.77}                                                        & 1404.00                                                        \\ \hline
\textsf{CMS}             & \multicolumn{1}{c|}{0.04}                                                                        & \multicolumn{1}{c|}{2.89}                                                            & 0.58                                                             & \multicolumn{1}{c|}{0.25}                                                             & 0.59                                                             & \multicolumn{1}{r|}{1766.67}                                                           & \multicolumn{1}{r|}{651.01}                                                        & 1399.92                                                        \\ \hline
\textsf{EMED}            & \multicolumn{1}{c|}{0.05}                                                                        & \multicolumn{1}{c|}{2.55}                                                            & 0.61                                                             & \multicolumn{1}{c|}{0.21}                                                             & 0.67                                                             & \multicolumn{1}{r|}{3526.78}                                                           & \multicolumn{1}{r|}{2062.03}                                                        & 7890.48                                                        \\  \hline 
\textsf{Bank}             & \multicolumn{1}{c|}{0.04}                                                                        & \multicolumn{1}{c|}{6.25}                                                            &                                                              0.60& \multicolumn{1}{c|}{0.26}                                                             &                                                             0.69& \multicolumn{1}{r|}{62.83}                                                           & \multicolumn{1}{r|}{21.55}                                                        &                                                       196.20\\ \hline
\textsf{IMSA}          & \multicolumn{1}{c|}{0.04}                                                                        & \multicolumn{1}{c|}{3.89}                                                            &                                                             0.53& \multicolumn{1}{c|}{0.25}                                                             &                                                             0.64& \multicolumn{1}{r|}{146.70}                                                           & \multicolumn{1}{r|}{52.82}                                                        &                                                         391.20\\ 
\bottomrule
\end{tabular}}
\end{center}
\end{table}

Notably, the retrieval times in Table~\ref{tab7} correspond to vector-based KG triple retrieval over Wikidata5M, whereas remote full-Wikidata and local Wikidata5M BFS latencies are reported separately in Appendix~\ref{sec:graph-retrieval-deployment-efficiency}.

\subsection{Ablation Study and Analysis}
\label{sec:ablation}
To understand the contributions of each module in our proposed KG-RAG4SM method, we examine the effectiveness and contributions of each module in the ablation study. 

\subsubsection{Abl-1: Subgraph Retrieval} \textbf{\textit{ Which subgraph retrieval techniques work better for KG-RAG4SM in terms of both effectiveness and efficiency to generate high-quality results for schema matching?}}

\begin{figure}[h]
\centering
\includegraphics[width=0.55 \textwidth]{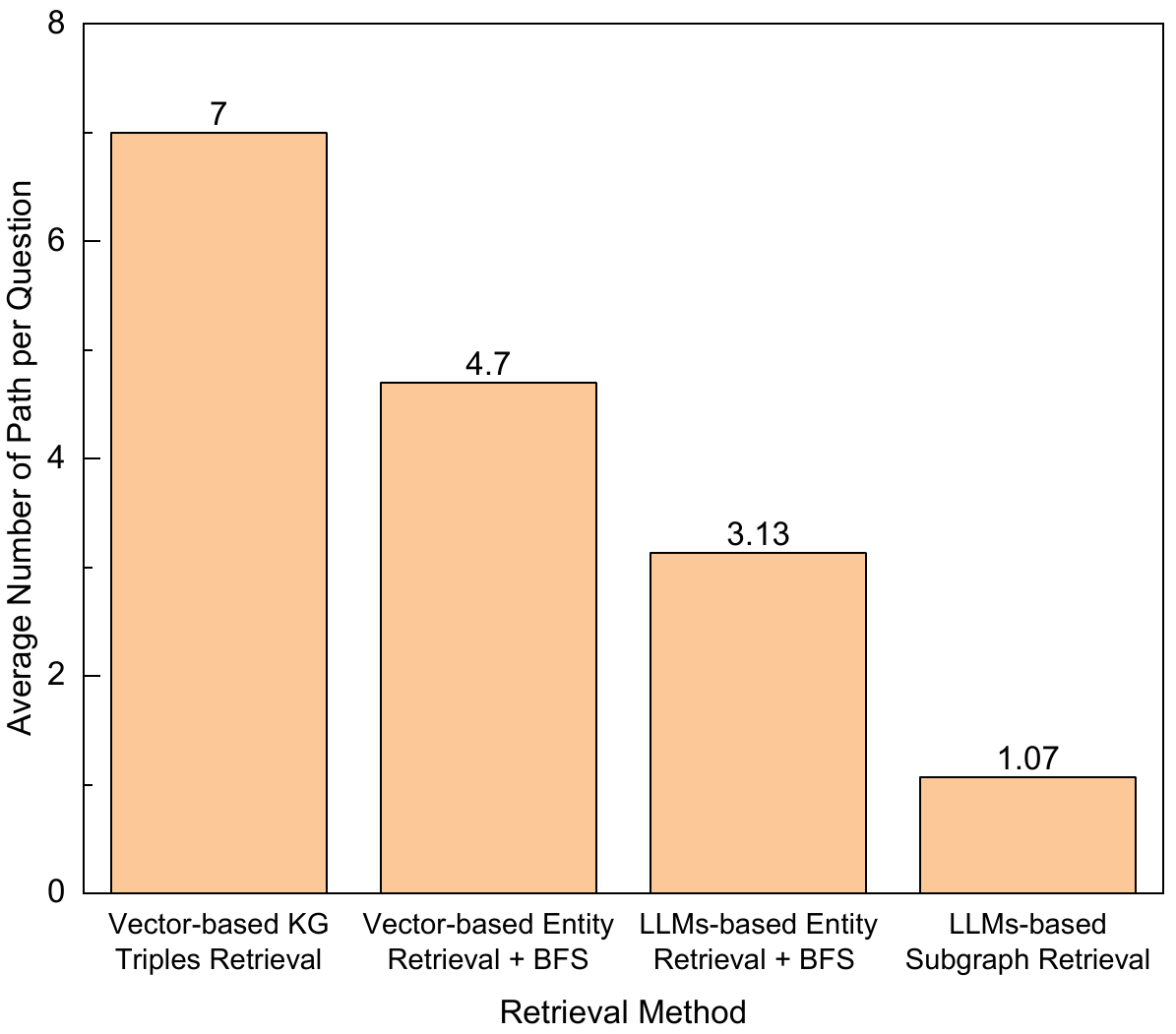}
\caption{Average number of retrieved subgraphs for different subgraph retrieval methods on the \textsf{CMS} dataset.} \label{fig:path_number}
\end{figure}

\noindent \underline{\textit{Evaluation Setting.}} We evaluate the performance of KG-RAG4SM with the retrieved subgraphs from different subgraph retrieval and refinement techniques. 
\textbf{(1) Vector-based retrieval.} We consider vector-based KG triple retrieval from Wikidata5M and vector-based entity retrieval using Wikidata5M embeddings followed by BFS over full-scale Wikidata.
In the W/O Rank setting (without ranking-based subgraph refinement), we retain up to the top-$10$ retrieved candidate subgraphs, whereas the ranking-based refinement settings select the top-$1$ and top-$2$ subgraphs based on our ranking scheme.
This is because retrieving and incorporating the full candidate triples is both computationally expensive and empirically misleading to LLMs and further substantially hurts the performance of the RAG system~\citep{yang2024kg}.
\textbf{(2) LLM-based retrieval.} We consider LLM-generated entity candidates followed by BFS over full-scale Wikidata and direct LLM-based subgraph retrieval.
There are two specific LLM-based subgraph retrieval methods: LLM-based entity retrieval combined with BFS-based graph traversal and LLM-based subgraph retrieval. 
Given that LLM-based retrieval directly prompts the LLM to identify only a small number of entities that actually exist in the Wikidata KG, yielding a very small number of candidate subgraphs, as shown in Figure~\ref{fig:path_number}, we evaluate it against vector-based retrieval without ranking-based subgraph refinement under the evaluated configurations.

\begin{table}
\caption{Effectiveness of KG-RAG4SM with GPT-4o-mini as the backbone LLM and different retrieved subgraphs on the \textsf{CMS} dataset without ranking-based subgraph refinement (W/O Rank).}
\label{tab8}
\small
\begin{center}
\begin{tabular}{c|cc||cc}
\toprule
\multirow{2}{*}{Metric (\%)}    & \multicolumn{2}{c||}{Vector-based Retrieval}                                                                                                                                                                               & \multicolumn{2}{c}{LLM-based Retrieval}                                                                                                                                                                          \\ \cline{2-5} 
                           & \multicolumn{1}{c|}{KG Triples Retrieval} & Entity Retrieval+BFS & \multicolumn{1}{c|}{Entity Retrieval+BFS} & Subgraph Retrieval \\ \midrule
\texttt{P}  & \multicolumn{1}{c|}{\underline{55.55}}                                                                                            & 47.05                                                                                             & \multicolumn{1}{c|}{\textbf{57.14}}                                                                                           & 44.44                                                                                       \\ \hline
\texttt{R}  & \multicolumn{1}{c|}{20.00}                                                                                            & \underline{32.00}                                                                                             & \multicolumn{1}{c|}{\textbf{48.00}}                                                                                           & \textbf{48.00}                                                                                       \\ \hline
\texttt{F1} & \multicolumn{1}{c|}{29.41}                                                                                            & 38.09                                                                                             & \multicolumn{1}{c|}{\textbf{52.17}}                                                                                           & \underline{46.15}                                                                                       \\ \bottomrule
\end{tabular}
\end{center}
\end{table}

\noindent \underline{\textit{Results Analysis.}} The results are given in Table \ref{tab8}, from which we find that under the \textsf{W/O Rank} setting, KG-RAG4SM with the subgraph retrieved using vector-based KG triples retrieval achieves 55.55\% \texttt{P} and 29.41\% \texttt{F1} while the subgraph retrieved using vector-based entity retrieval+BFS achieves 47.05\% \texttt{P} and 38.09\% \texttt{F1}.  
We also observe that the KG-RAG4SM with the subgraph retrieved via LLM-based entity retrieval + BFS achieves 57.14\% \texttt{P} and 52.17\% \texttt{F1}, while the subgraph retrieved via LLM-based subgraph retrieval achieves 44.44\% \texttt{P} and 46.15\% \texttt{F1}.  
This suggests that on the \textsf{CMS} dataset, KG-RAG4SM with GPT-4o-mini as the backbone LLM and with the subgraph retrieved by using vector-based KG triples retrieval outperforms the vector-based entity retrieval + BFS in terms of \texttt{P} but underperforms in terms of \texttt{F1} score.
Meanwhile, KG-RAG4SM with the subgraph retrieved by using LLM-based entity retrieval + BFS outperforms LLM-based subgraph retrieval in terms of \texttt{P} and \texttt{F1} on the \textsf{CMS} dataset. 
In terms of the efficiency of the subgraph retrieval, the average number of retrieved paths per question for these subgraph retrieval strategies on the \textsf{CMS} dataset is shown in Figure \ref{fig:path_number}. 
We observe that the average number of retrieved subgraphs per question is 7 and 4.7 using vector-based KG triples retrieval and vector-based entity retrieval+BFS, while it is 3.13 and 1.07 using LLM-based entity retrieval+BFS and LLM-based subgraph retrieval. 
This suggests that vector-based retrieval can retrieve more subgraphs than LLM-based retrieval because LLMs may have limited reasoning and understanding capabilities about commonsense knowledge graphs (e.g., Wikidata).
As a result, we cannot retrieve any subgraphs for the majority of questions by using the LLM-based subgraph retrieval method. 
The average retrieval time per question for vector-based KG triples retrieval and vector-based entity retrieval + BFS is 2.88 seconds and 7.43 seconds, respectively, while the retrieval time per question for LLM-based Entity Retrieval + BFS and LLM-based subgraph retrieval is 1.40 seconds and 0.66 seconds, respectively. Even though LLM-based retrieval is faster than vector-based retrieval, due to the higher number of retrieved subgraphs, vector-based retrieval is preferred.

\begin{table}
\caption{Effectiveness of KG-RAG4SM with GPT-4o-mini using Wikidata5M vector-based KG triple retrieval and hybrid entity retrieval with BFS over full-scale Wikidata, with and without ranking-based refinement, on the \textsf{CMS} dataset.}
\label{tab9}
\small
\begin{center}
\begin{tabular}{c|ccc||ccc}
\toprule
\multicolumn{1}{c|}{\multirow{2}{*}{Metric (\%)}} & \multicolumn{3}{c||}{Vector-based KG Triples Retrieval}     & \multicolumn{3}{c}{Vector-based Entity Retrieval+BFS}                                                                                                              \\ \cline{2-7} 
\multicolumn{1}{c|}{}                        & \multicolumn{1}{c|}{Top-1 Ranked} & \multicolumn{1}{c|}{Top-2 Ranked} & W/O Rank & \multicolumn{1}{c|}{Top-1 Ranked} & \multicolumn{1}{c|}{Top-2 Ranked} & W/O Rank \\ \midrule
\texttt{P}                                    & \multicolumn{1}{c|}{\textbf{62.50}}                                         & \multicolumn{1}{c|}{52.38}                                                  & \multicolumn{1}{c||}{\underline{55.55}}                         & \multicolumn{1}{c|}{\textbf{59.09}}                                                  & \multicolumn{1}{c|}{\underline{52.63}}                                                  & 47.05                                              \\ \hline
\texttt{R}                                    & \multicolumn{1}{c|}{\underline{40.00}}                                                  & \multicolumn{1}{c|}{\textbf{44.00}}                                                  & 20.00                                              & \multicolumn{1}{c|}{\textbf{52.00}}                                         & \multicolumn{1}{c|}{\underline{40.00}}                                                  & 32.00                                              \\ \hline
\texttt{F1}                                   & \multicolumn{1}{c|}{\textbf{48.70}}                                                  & \multicolumn{1}{c|}{\underline{47.82}}                                                  & 29.41                                              & \multicolumn{1}{c|}{\textbf{55.31}}                                         & \multicolumn{1}{c|}{\underline{45.45}}                                                  & 38.09                                              \\ \bottomrule
\end{tabular}   
\end{center}
\end{table}

Next, we evaluate the effectiveness of the ranking-based subgraph refinement for the vector-based retrieval strategy. The final results with different ranked subgraphs and the subgraphs without ranking on the \textsf{CMS} dataset are reported in Table \ref{tab9}. 
We observe that the model augmented with the top-2 ranked subgraphs retrieved by vector-based KG triples retrieval contributes to 120\% and 62.59\% improvements on \texttt{R} and \texttt{F1} score, respectively, compared with the model without ranking. 
Meanwhile, the top-1 ranked subgraphs retrieved by vector-based entity retrieval + BFS contribute to 25.59\%, 62.50\%, and 45.20\% improvements on \texttt{P}, \texttt{R}, and \texttt{F1} score, respectively, compared with the model without (W/O) rank.
Overall, the results suggest that the model augmented with ranking-based refinement outperforms the model augmented without ranking.
This is because our ranking scheme can identify the most relevant knowledge by removing the irrelevant entities, relations, and KG triples from the retrieved subgraphs.

It is worth noting that as shown in Table~\ref{tab9}, with top-1 ranked paths, vector-based entity retrieval followed by BFS achieves an \texttt{F1} score of 55.31\%, compared with 48.70\% for vector-based KG triple retrieval. 
This performance gap may also result from the different configurations in both retrieval strategy and KG coverage: the former uses Wikidata5M embeddings to retrieve seed entities and then discovers multi-hop paths over the full-scale Wikidata KG via SPARQL-based BFS, whereas the latter retrieves triples directly from Wikidata5M. 
The observed gap therefore cannot be attributed to either factor alone, and Table~\ref{tab9} should be read as a comparison of the two configurations rather than an isolated assessment of retrieval strategy or KG coverage.

\subsubsection{Abl-2: Backbone LLMs}
\textbf{\textit{Which LLMs perform better in KG-RAG4SM, e.g., do fewer-parameter LLMs work better than large-size LLMs for KG-RAG4SM?}}

\noindent \underline{\textit{Evaluation Setting.}}  We evaluate the performance of KG-RAG4SM with the mainstream backbone LLMs varying from small to large sizes. 
GPT-4o-mini, GPT-4o, and Mistral-7B-Instruct from OpenAI and Mistral AI are selected.
Fine-tuned models, such as Jellyfish-7B and Jellyfish-8B, are also considered as the backbone LLMs from~\citep{zhang2024jellyfish}.
These models are fine-tuned based on Mistral-7B-Instruct and Llama-3-8B-Instruct, respectively, using a subset of instruction data for error detection, data imputation, schema matching, and entity matching.
We report the results of our proposed KG-RAG4SM on the \textsf{CMS} dataset with the top-$2$ ranked subgraphs retrieved via the vector-based KG triples retrieval method in Table \ref{tab10}.

\begin{table}[ht!]
\caption{Results of KG-RAG4SM with different backbone LLMs and the top-2 ranked subgraphs using vector-based KG triples retrieval on the \textsf{CMS} dataset.} 
\label{tab10}
\small
\begin{center}
\begin{tabular}{c|c|c|c|c|c}
\toprule
\multirow{2}{*}{Metric (\%)} & \multicolumn{2}{c|}{GPT}     & \multicolumn{2}{c|}{Jellyfish}                                                   & \multicolumn{1}{c}{Mistral}                                                  \\ \cline{2-6} 
                  & \multicolumn{1}{c|}{4o-mini} & \multicolumn{1}{c|}{4o} & \multicolumn{1}{c|}{8B} & \multicolumn{1}{c|}{7B} & \multicolumn{1}{c}{7B} \\ \midrule
\texttt{P}                 & \multicolumn{1}{c|}{\underline{52.38}}            & \multicolumn{1}{c|}{34.78}       &  23.33             & \multicolumn{1}{c|}{\textbf{66.66}}              &      26.66                              \\ \hline
\texttt{R}                 & \multicolumn{1}{c|}{\underline{44.00}}            & \multicolumn{1}{c|}{32.00}       &  28.00             & \multicolumn{1}{c|}{16.00}              &  \textbf{64.00}                                  \\ \hline
\texttt{F1}                & \multicolumn{1}{c|}{\textbf{47.82}}           & \multicolumn{1}{c|}{33.33}       & 25.45              & \multicolumn{1}{c|}{25.81}              &        \underline{37.64}                            \\ \hline
Time (Sec)  & \multicolumn{1}{c|}{\underline{0.58}}            & \multicolumn{1}{c|}{0.72}       &\textbf{0.39}               & \multicolumn{1}{c|}{0.64}              & 5.62                                    \\ \bottomrule
\end{tabular}
\end{center}
\end{table}

\noindent \underline{\textit{Results Analysis.}} We find that KG-RAG4SM with GPT-4o-mini as the backbone LLM attains \textbf{52.38\%}, 44.00\%, and \textbf{47.82\%} \texttt{P}, \texttt{R}, and \texttt{F1} score, respectively, while with GPT-4o as the backbone LLM, it achieves 34.78\%, 32.00\%, and 33.33\% \texttt{P}, \texttt{R}, and \texttt{F1} score, respectively. 
In terms of generation times, KG-RAG4SM with GPT-4o-mini is faster than that with GPT-4o. This indicates that KG-RAG4SM works better with GPT-4o-mini than with GPT-4o. Generally, KG-RAG4SM with small-sized LLMs works better than with large-sized ones.
This is due to the fact that frequent knowledge conflicts might occur between the retrieved knowledge and the internal knowledge of large-sized LLMs.
We also observe that KG-RAG4SM with Jellyfish-7B as the backbone LLM achieves \textbf{66.66\%} \texttt{P} in 0.64 seconds per question, while with Mistral-7B-Instruct it attains \textbf{26.66\%} \texttt{P} in 5.62 seconds per question. This suggests the effectiveness of the Jellyfish model in KG-RAG4SM since Jellyfish-7B is a fine-tuned model based on Mistral-7B-Instruct.  
Overall, we find that the choice of backbone LLMs is crucial for KG-RAG4SM, where the small-sized LLMs work better than the large-sized ones.

\subsection{Extended Experimental Results}
To further analyze the scalability and truthfulness of KG-RAG4SM, we run extended experiments with domain-specific KGs and study the effectiveness of KG-RAG4SM with the retrieved poorly relevant KG context.
\subsubsection{Ext-1: Scalability across KG Domains}
\textbf{\textit{Which KGs work better in KG-RAG4SM, e.g., do domain-specific KGs work better than commonsense KGs for KG-RAG4SM?}}

\noindent \underline{\textit{Evaluation Setting.}} To demonstrate the scalability across KG domains of KG-RAG4SM,  we extend the domain of KGs from commonsense KGs (Wikidata) to domain-specific KGs (Medical KGs) and evaluate the effectiveness of medical KGs in KG-RAG4SM. 
In view of our schema matching dataset originating from the healthcare databases, we retrieve and rank the relevant entities from the UMLS KG based on vector similarity and then expand each entity with retrieved relations from the UMLS KG by using the UMLS API\footnote{https://documentation.uts.nlm.nih.gov/rest/home.html} to form the triples. 
Specifically, we leverage UMLS-BERT~\citep{michalopoulos2021umlsbert} to encode the given schema matching question and then query the relevant entities from UMLS, and next retrieve the relation for each entity by using the caching mechanism to accelerate relation retrieval from UMLS API endpoints.
For each entity, we expand it with queried relations as triples and then rank the triples based on the similarity score between the question and the entity. 
We present results of KG-RAG4SM with GPT-4o-mini and Jellyfish-8B as backbone LLMs and the retrieved \& ranked triples from UMLS KG in comparison to the triples from Wikidata by using vector-based KG triple retrieval on the \textsf{Synthea} dataset in Table \ref{tab11}.

\begin{table}
\caption{Comparison of KG-RAG4SM with GPT-4o-mini \& Jellyfish-8B as the backbone LLM and different KGs on \textsf{Synthea} dataset.} \label{tab11}
\small
\begin{center}
\begin{tabular}{c|c|c|c|c||c|c|c|c}
\toprule
\multirow{3}{*}{Metric (\%)}  & \multicolumn{4}{c||}{GPT-4o-mini} & \multicolumn{4}{c}{Jellyfish-8B} \\
\cline{2-9} 
& \multicolumn{2}{c|}{UMLS KG} & \multicolumn{2}{c||}{Wikidata KG} & \multicolumn{2}{c|}{UMLS KG} & \multicolumn{2}{c}{Wikidata KG} \\ 
\cline{2-9} 
                           & \begin{tabular}[x]{@{}c@{}}Top-2\\Ranked\end{tabular} & \begin{tabular}[x]{@{}c@{}}Top-1\\Ranked\end{tabular} & \begin{tabular}[x]{@{}c@{}}Top-2\\Ranked\end{tabular}& \begin{tabular}[x]{@{}c@{}}Top-1\\Ranked\end{tabular}  & \begin{tabular}[x]{@{}c@{}}Top-2\\Ranked\end{tabular} & \begin{tabular}[x]{@{}c@{}}Top-1\\Ranked\end{tabular} & \begin{tabular}[x]{@{}c@{}}Top-2\\Ranked\end{tabular} & \begin{tabular}[x]{@{}c@{}}Top-1\\Ranked\end{tabular} \\ 
\midrule
\texttt{P}                  & 11.11 & \underline{16.67} & \textbf{20.00} & 15.00   & 20.00 & 18.18 & \textbf{26.31} & \underline{25.00} \\ 
\hline
\texttt{R}                 & 9.09  & 18.18 & \textbf{36.36} & \underline{27.27} & \underline{36.36}  & \underline{36.36} & \textbf{45.45} & \textbf{45.45} \\  
\hline
\texttt{F1}                & 10.00 & 17.39 & \textbf{25.80} & \underline{19.35}  & 25.81 & 24.24 & \textbf{33.33} & \underline{32.25} \\  
\bottomrule
\end{tabular}
\end{center}
\end{table}

\noindent \underline{\textit{Results Analysis.}} We find that KG-RAG4SM with UMLS KG achieves comparable yet worse results in comparison with the results of KG-RAG4SM having triples from Wikidata KG across the metrics on the \textsf{Synthea} dataset, both with GPT-4o-mini and Jellyfish-8B as backbone LLMs. 
The reasons are twofold: \textbf{(1)} {\em Relevance and Quality of KG:} Although all of our schema matching datasets originate from healthcare databases, the questions of schema matching pairs, including table name (\texttt{admission, patient, visit, insurance, service}, etc.) and attribute name (\texttt{admission\_location, patient\_weight, visit\_date, insurance\_id, service\_provider}, etc.) from \texttt{source} and \texttt{target data models,} are more relevant to the entities and taxonomic relations (e.g., \texttt{patient, instance of, human}, etc.) from commonsense KGs than medical entities and relations (e.g., \texttt{diabetes status: Prid: Pt: $\hat{}$ Patient: Nom, has\_supersystem, patient}, etc.) from medical KGs.
Most of the triples that exist in UMLS KG are one-hop relations, as the UMLS maintains the entities and their associated one-hop relations, while the multiple-hop relations are absent~\citep{yang2024kg}. \textbf{(2)} {\em Answers Prone to False Positives:} Similar to other RAG systems, Graph RAG and KG-RAG systems can be easily distracted by irrelevant context~\citep{songmeasuring2025} and are prone to generating false positive answers with ambiguous or poorly relevant KG context. 
In particular, the results get worse when it comes to the lower relevance and quality of the retrieved KG triples, such as triples retrieved from the UMLS KG in our schema matching case. 
For instance, the KG triples from the UMLS KG  show the connections between \texttt{stopping medication} and \texttt{drug exposure}, and the LLMs incorrectly infer that the reason for \texttt{stopping medication} is associated with \texttt{drug exposure}. However, the retrieved connections explicitly mention that \texttt{stopping medication (120234)} is a synonym of \texttt{Medication discontinued (274512008)}, which is associated with \texttt{Drug therapy (416608005)}, but \texttt{Drug therapy (416608005)} is a subclass of \texttt{Drug or medicament (410942007)} rather than a semantically equivalent relation. 
Since LLMs are prone to generating false positive answers based on the found connections from the provided KG context, this further induces decreased performance of KG-RAG4SM in this scenario. 
Notice that we have not employed rule-based relevance scoring and instruction tuning (\S \ref{subs:instruction_tuning}) in this experiment, and we demonstrate this module's impact separately later in \S \ref{sec:robust_exp}. 

\subsubsection{Ext-2: Truthfulness of KG-RAG4SM} \label{sec:robust_exp}  \textbf{\textit{Does the rule-based relevance scoring and instruction-tuning work for improving the truthfulness of KG-RAG4SM?}}

\noindent \underline{\textit{Evaluation Setting.}} To mitigate the truthfulness issue of KG-RAG4SM with poorly relevant and lower-quality KG triples from the UMLS KG, we further introduce the rule-based relevance scoring with instruction tuning (\S \ref{subs:instruction_tuning}).  
Before feeding the retrieved triples from the KG to LLMs, we call LLMs to assign a relevance score between the provided KG context and the given schema matching questions with few-shot examples and rule-based instructions. 
According to the rule-based instruction, LLMs can generate the answer based on their own knowledge when the relevance score between the provided KG context and the given schema-matching questions is lower than 7, or the KG context is unavailable. 
We evaluate the effectiveness of rule-based relevance scoring with instruction tuning in KG-RAG4SM using Jellyfish-8B as the backbone LLM on the \textsf{Synthea} dataset.

\noindent \underline{\textit{Results Analysis.}} Table~\ref{tab13} shows the results of KG-RAG4SM with rule-based relevance scoring and instruction-tuning in comparison to KG-RAG4SM without instruction-tuning, with the settings of leveraging Jellyfish-8B as the backbone LLM and UMLS KGs as domain-specific KGs on the \textsf{Synthea} dataset. 
We observe that KG-RAG4SM with rule-based relevance scoring and instruction tuning contributes significant improvements across metrics on the \textsf{Synthea} dataset with Jellyfish-8B as the backbone LLM in comparison to KG-RAG4SM without instruction tuning.
In particular, the KG-RAG4SM with rule-based relevance scoring and instruction tuning and top-1 ranked triples from UMLS KG contributes to the \textbf{96.42\%}, \textbf{25.00}\%, and \textbf{65.01\%} improvement on \texttt{P}, \texttt{R}, and \texttt{F1} score, respectively, on the \textsf{Synthea} dataset with Jellyfish-8B as the backbone LLM in comparison to the results of KG-RAG4SM without instruction-tuning. These results are comparable and often better than the results of KG-RAG4SM having triples from the Wikidata KG on the \textsf{Synthea} dataset presented earlier in Table \ref{tab11}. 
This demonstrates the effectiveness and truthfulness of KG-RAG4SM with the designed rule-based relevance scoring and instruction-tuning in mitigating false-positive-prone answers and hallucinations of RAG systems with poorly relevant retrieved KG context.   

\begin{table}
\caption{Effectiveness of rule-based relevance scoring and instruction tuning (IT) in KG-RAG4SM with Jellyfish-8B and UMLS KGs on the \textsf{Synthea} dataset.} 
\label{tab13}
\small
\begin{center}
\begin{tabular}{c|c|c||c|c}
\toprule
\multirow{2}{*}{Metric (\%)} & \multicolumn{2}{c||}{KG-RAG4SM W/O Rule-based IT} & \multicolumn{2}{c}{KG-RAG4SM With Rule-based IT} \\ 
\cline{2-5} 
                                             & Top-2 Ranked & Top-1 Ranked & Top-2 Ranked & Top-1 Ranked \\ 
\midrule
\texttt{P}               & 20.00 & 18.18 & \underline{33.33}($\uparrow$66.65\%) & \textbf{35.71}($\uparrow$96.42\%) \\ 
\hline
\texttt{R}               & \underline{36.36}  & \underline{36.36} & \textbf{45.45}($\uparrow$25.00\%) & \textbf{45.45}($\uparrow$25.00\%) \\ 
\hline
\texttt{F1}             & 25.81 & 24.24 & \underline{38.46}($\uparrow$49.01\%) & \textbf{40.00}($\uparrow$65.01\%) \\ 
\bottomrule
\end{tabular}
\end{center}
\end{table}

\subsubsection{Ext-3: Ablation Study on Subgraph Refinement in KG-RAG4SM and Comparison with SOTA GraphRAG.}
We investigate the following two research questions: \textbf{\textit{(Ext-3a)  What is the isolated contribution of the similarity ranking-based subgraph refinement mechanism to the performance of KG-RAG4SM?
(Ext-3b) How does the complete KG-RAG4SM framework perform relative to the state-of-the-art GraphRAG baseline?}}

\noindent \underline{\textit{Evaluation Setting.}} \textbf{(Ext-3a):} To evaluate the contribution of the similarity ranking-based subgraph refinement module, we conduct a controlled internal ablation study. 
Specifically, we retain the retrieved paths up to the top-10 candidate triples by similarity score but do not further refine down to a single top-ranked path in the KG-RAG4SM W/O Rank setting. 
We then compare it to KG-RAG4SM (top-1 Ranked), which retains only the single highest-ranked path while holding all other settings constant, to isolate the effect of similarity ranking-based subgraph refinement.
\textbf{(Ext-3b):} We further examine the strengths of KG-RAG4SM in terms of overall performance and computational cost against the SOTA GraphRAG baseline. 
We select ArchRAG~\citep{WangFZLM26} as the baseline, since it has already demonstrated the best overall performance among existing GraphRAG approaches on question answering.
To make a fair comparison, we ran KG-RAG4SM and ArchRAG with the same embedding model, i.e., RoBERTa-base, and kept the same backbone LLM (GPT-4o-mini), both for ArchRAG's community summarization and inference.
It is worth mentioning that the mainstream GraphRAG approach, like ArchRAG, is intentionally designed to retrieve the subgraph from the self-built small-sized KG. Therefore, it is challenging to directly employ it to large-sized KGs. 
To adapt ArchRAG to a large-sized Wikidata5M KG on the schema matching task, we call an LLM to augment the graph via hierarchical clustering and community summary only for the top-$15$ entities that are connected to the retrieved entity for a specific query, instead of all entities in Wikidata5M. 
The reasons are twofold: (1) LLM-based hierarchical clustering and community summarization over the full-scale KG, comprising billions of entities, would incur substantially higher token costs \footnote{ArchRAG, in its original implementation, will use LLMs to augment the graph through hierarchical clustering and community summarization on Wikidata5M (4.59 million entities). This process would generate 0.732 million community reports and level summaries, consuming 1.4 billion tokens and incurring an amortized cost of approximately \$400 with gpt‑4o‑mini, solely for clustering and summarization.}; (2) Large communities clustered over the full-scale KG produce larger contextual knowledge, which may mislead the LLM's inference and further degrade overall performance due to noisy knowledge. We keep other implementation details of ArchRAG the same as its original implementation.
We present the overall results, computation cost, and end‑to‑end latency of KG‑RAG4SM, which retrieves and ranks subgraphs from Wikidata5M via vector‑based KG triple retrieval, compared against ArchRAG that uses hierarchical search on the C‑HNSW index (highest layer $L=5$) to retrieve context from Wikidata5M KG. The comparison on the \textsc{CMS} and \textsf{BANK} datasets is reported in Table~\ref{tab:abl_archrag}.

\begin{table}
\caption{(Ext-3a) Ablation study on similarity ranking-based subgraph refinement in KG-RAG4SM (Top-1 Ranked vs. W/O ranking) with GPT-4o-mini; and (Ext-3b) comparison against SOTA GraphRAG baseline ArchRAG on \textsf{CMS} and \textsf{BANK} dataset. } 
\label{tab:abl_archrag}
\small
\begin{center}
\small
\begin{tabular}{c|c|c|c||c|c|c}
\toprule
\multirow{3}{*}{Metric}& \multicolumn{3}{c||}{\textsf{CMS}}& \multicolumn{3}{c}{\textsf{Bank}}\\ \cline{2-7}
 & \multicolumn{2}{c|}{Controlled Ablation}& Baseline & \multicolumn{2}{c|}{Controlled Ablation}& Baseline\\ 
\cline{2-7}& \begin{tabular}[x]{@{}c@{}}KG-RAG4SM \\Top-1 Ranked\end{tabular} & \begin{tabular}[x]{@{}c@{}}KG-RAG4SM \\W/O Rank\end{tabular} & ArchRAG & \begin{tabular}[x]{@{}c@{}}KG-RAG4SM \\Top-1 Ranked\end{tabular} & \begin{tabular}[x]{@{}c@{}}KG-RAG4SM \\W/O Rank\end{tabular} & ArchRAG\\ 
\midrule
\texttt{P} (\%)               & \textbf{62.50}& 55.55& \underline{46.15} & \textbf{33.33}& \textbf{33.33}&\underline{28.57}\\ 
\hline
\texttt{R} (\%)               & \textbf{40.00}& 20.00& \underline{24.00} & \textbf{100}& \textbf{100}&\textbf{100}\\ 
\hline
\texttt{F1} (\%)             & \textbf{48.70}& 29.41& \underline{31.58} & \textbf{50.00}& \textbf{50.00}&\underline{44.44}\\ \hline
 \# API Calls & \textbf{2,563} & \textbf{2,563} & \underline{3,379} & \textbf{146}& \textbf{146}&\underline{147}\\ \hline
 Tokens (M) & \textbf{2.57} & \underline{3.00} &    4.43 & \textbf{0.14}& 0.17&\underline{0.16}\\  \hline
 Latency (Sec) & \textbf{3.75}& \underline{3.92} &   7.98 & \textbf{7.06}& \underline{7.08}&17.34\\
  \bottomrule
\end{tabular}
\end{center}
\end{table}

\noindent \underline{\textit{Results Analysis.}} 
As shown in Table~\ref{tab:abl_archrag}, we evaluate KG-RAG4SM through a controlled internal ablation isolating the ranking-based subgraph refinement, alongside an overall framework comparison against the SOTA baseline ArchRAG. (1) {\bf Ext-3a:} KG-RAG4SM with top-1 ranked triples retrieved from Wikidata5M outperforms KG-RAG4SM W/O Rank that retains up to top-10 candidates via similarity score without further rank-based refinement, in \texttt{F1} score, token consumption, and latency on \textsf{CMS}, while achieving comparable \texttt{F1} score with lower token consumption and latency on \texttt{Bank}. These ablation results empirically demonstrate that similarity ranking-based subgraph refinement effectively prunes irrelevant triples from the retrieved candidates, mitigating context noise and contributing to both the matching quality and cost efficiency of KG-RAG4SM.
(2) {\bf Ext-3b:} Compared to the SOTA GraphRAG baseline ArchRAG, KG-RAG4SM with top-1 ranked triples retrieved from Wikidata5M consistently outperforms ArchRAG across all effectiveness metrics (except for \texttt{R} on \textsf{Bank} where both methods achieve a 100\% \texttt{R}), while significantly reducing the number of LLM API calls, token consumption, and end-to-end latency on \textsf{CMS}. Specifically, KG-RAG4SM with top-1 ranked triples achieves the best \texttt{F1} score on both datasets while maintaining minimal token consumption and end-to-end latency relative to ArchRAG. This is primarily attributable to the significantly higher number of LLM calls and token consumption required by ArchRAG's hierarchical clustering and community summarization, despite applying the aforementioned optimization to accommodate the large-scale Wikidata KG.

These results confirm that KG-RAG4SM, leveraging retrieved and ranked subgraphs via vector-based KG triple retrieval, excels in both performance and efficiency for schema matching compared to the ArchRAG baseline when retrieving from large external KGs. This further highlights the robustness and adaptivity of KG-RAG4SM in real-world data integration scenarios where internal contextual knowledge is unavailable.

\subsection{Case Study} \label{susec:case_study}
A case study is conducted to further verify the hallucination mitigation of our KG-RAG4SM method on the complex schema matching case from real-world scenarios.

\subsubsection{Case Introduction}
We select schema matching tasks in a real-world scenario by mapping the data tables from the e-MedSolution\footnote{https://www.eszfk.hu/e-medsolution} clinical module into the OMOP CDM data model. 
The clinical module of e-MedSolution is selected to map into the OMOP CDM clinical data model, since the concise schema is the core and representative module of the e-MedSolution system and OMOP CDM is a standardized healthcare data model~\citep{jiang2017consensus}.
Thereby, a new schema-level schema matching dataset, \textsf{EMED}, is created by referring to the format of the \textsf{OMAP} dataset.
It is worth mentioning that only the schema-level data, i.e., entities, attributes, and constraints, of the e-MedSolution are accessed when creating the \textsf{EMED} dataset.

\subsubsection{Case: Hallucinations}
To illustrate hallucination mitigation of KG-RAG4SM versus LLM-based schema matching, we conduct a case study to demonstrate the roles of the retrieved subgraph $G$ from the large-scale $\mathcal{KG}$. 

\noindent \underline{\textit{Evaluation Setting.}} 
We set up the case study with the following subgraph retrieval and backbone LLMs. \textbf{(1)} Subgraph retrieval: the refined top-$1$ subgraphs by using vector-based $\mathcal{KG}$ triples retrieval. \textbf{(2)} Backbone LLMs: GPT-4o-mini. 
To highlight the hallucination mitigation of our KG-RAG4SM, we ask the LLM to return the results for the given schema-matching questions and provide explanations.

\begin{figure}[h]
\centering
\includegraphics[width=1.0 \textwidth]{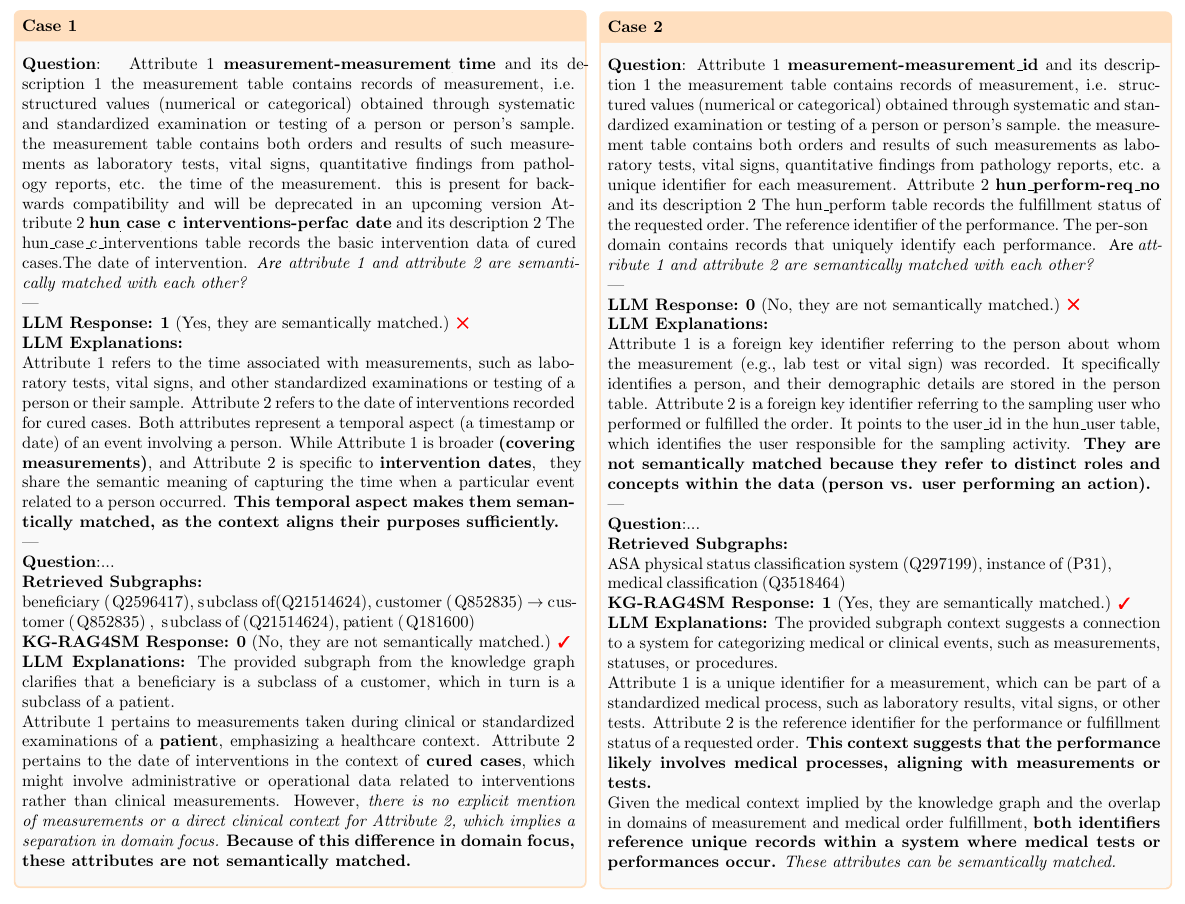}
\caption{Case study: Hallucination mitigation of KG-RAG4SM for schema matching. The schema matching question for KG-RAG4SM is not shown, since it is the same question given to the LLM.  } \label{figcase}
\end{figure}

\noindent \underline{\textit{Case Verification.}} 
The two schema matching cases are selected from the \textsf{EMED} dataset.  
Figure \ref{figcase} demonstrates the responses of the KG-RAG4SM model and its correctness compared with ground-truth answers by highlighting how KG-RAG4SM mitigates hallucinations with the retrieved subgraphs from large-scale $\mathcal{KG}$. 
The \textcolor{red}{\footnotesize\CheckmarkBold} and \textcolor{red}{\footnotesize\XSolid} indicate the correctness of the responses compared to ground truth.

\noindent \underline{\textit{Result Analysis.}} In the first case, the response of the LLM is \textbf{1}, indicating the given attributes are semantically matched, while these attributes are not semantically matched in both the ground-truth answer and schema matching in practice since the \texttt{measurement-measurement\_time} and \texttt{hun\_case\_c\_interventions-perfac\_date} are different from each other. 
The reason behind this response is as follows: LLM incorrectly thought that both attributes represent a temporal aspect of an event involving a person, and this temporal aspect makes them semantically matched. 
In contrast, KG-RAG4SM, augmented with the relevant subgraphs retrieved from the external $\mathcal{KG}$, responds with the answer \textbf{0}, indicating that the given attributes are not semantically matched.
This is mainly because the retrieved subgraph \texttt{beneficiary (Q2596417), subclass of (Q21514624), customer (Q852835) $\rightarrow$ customer (Q852835), subclass of (Q21514624), patient (Q181600)} clarify the hierarchical relationships among \texttt{beneficiary}, \texttt{customer}, and \texttt{patient}, which helps LLMs to establish the hierarchical domain relationships of the attributes in the given schema matching question.
Even if the retrieved subgraph does not explicitly mention \texttt{measurements} or a direct clinical context for \texttt{interventions}, it implies that there is a separation in domain focus with different contexts.
This strengthens the conclusion that they are not semantically matched, which is consistent with the response of KG-RAG4SM and the ground-truth answer.

In the second case, the response of the LLM is \textbf{0} while these attributes are semantically matched in the ground truth. 
Interestingly, the LLM incorrectly thinks that these two attributes refer to distinct roles and concepts between a person and a user performing a medical action.
In contrast, KG-RAG4SM returns the response with the correct answer \textbf{1} with the help of the retrieved subgraph from $\mathcal{KG}$: \texttt{ASA physical status classification system (Q297199), instance of (P31), and medical classification (Q3518464)}.
The provided subgraph suggests a connection to a system for categorizing medical or clinical events, such as measurements, statuses, or procedures, and the performance likely involves medical processes, aligning with measurements or tests.
It implicitly shows the overlap in domains of measurement and medical order fulfillment, which indicates that both identifiers refer to unique records within a system where medical tests or performances occur.

The two case studies demonstrate that the hallucination issues of LLMs for schema matching are well mitigated by augmenting the LLMs with the retrieved relevant subgraphs from large-size knowledge graphs in our KG-RAG4SM. 

\section{Related Work} 
\label{sec5}
\noindent \textbf{Schema Matching and Data Integration.}
Schema matching is a process of establishing semantic associations between different schemata, in which the semantic correspondences between the source and target schemas are usually identified~\citep{Rahm2011}.
In contrast to schema matching, schema mapping is a process of generating the assertions and mappings from the identified semantic correspondence, by which the source schema can be mapped onto the target schema to provide a common interface for accessing and querying heterogeneous data~\citep{Fuxman2009}. 
In general, the correspondences between the source and target schemas can be identified manually by domain experts, which requires a lot of human effort and experience~\citep{ma2023thesis}. 
In the case of manual mapping, mismatches and errors are also inevitable, as the intensive and repeated work might distract humans.  
To improve the matching efficiency and to minimize mismatches, an auto-match system is designed to support automatic schema matching based on machine learning~\citep{berlin2002database}. 
In this method, probability knowledge is obtained from domain experts based on Bayesian learning, which provides an attribute dictionary to find optimal matching pairs. 
Considering the strengths of similarity-based matching and machine learning-based matching, a hybrid schema matching model is proposed by combining lexical and semantic similarity with machine learning~\citep{bulygin2018}.

To free human hands from the tedious schema matching task, a new semi-automatic schema matching approach is proposed~\citep{drumm2007quickmig}.
In this approach, the domain ontologies and sample instances are reused to identify and determine the semantic correspondences between schemata. 
Similarly, background ontologies are utilized to provide the semantic reference for schema matching based on ontology alignment~\citep{nathalie2009}.
To minimize mismatches and errors, the medical knowledge bases from the medical encyclopedia are utilized to learn better representations of medical concepts and properties.
Thereby, a medical schema matching method is proposed based on the relation alignment between the description embedding of an extracted medical entity and learned heterogeneous knowledge graph embedding~\citep{gaowjw22}.
However, these methods follow supervised learning, in which large labeled data are required for model training. 

Recently, several schema matching approaches based on LLMs have been proposed to address the above limitations, such as fine-tuned LLMs Magneto~\citep{LiuMagneto25}, retrieval-enhanced LLMs~\citep{sheetrit2024rematch}, and reasoning-guided LLMs~\citep{xu2024kcmf}.
To improve the performance of LLMs for data preprocessing, an instruction-tuning LLM~\citep{zhang2024jellyfish} is proposed, which enables optional knowledge injection during inference through instruction tuning with reasoning based on domain knowledge. 
Since the inherent knowledge might be useful to enhance LLMs for schema matching, RAG-based approaches are proposed to augment LLMs by retrieving the relevant knowledge from text chunks constructed based on the textual descriptions of tables and attributes, e.g., a retrieval-enhanced large language model for schema matching~\citep{sheetrit2024rematch} that retrieves the top-$k$ documents from the textual documents based on semantic relevance.
Inspired by RAG on LLMs, a compositional language model for schema matching~\citep{seedat25bootstrapping} enhances and refines the results based on semantic retrieval from multi-vector documents and confidence scoring by multiple calls of LLMs.
Similarly, an LLM-based knowledge-compliant schema and entity-matching framework~\citep{xu2024kcmf} is proposed that guides LLM reasoning based on pseudo-code-based task decomposition and knowledge retrieval from domain knowledge sets by calling LLMs to query the relevant keywords.

The prevailing methods of schema matching could be classified into five categories: similarity-based matching~\citep{berlin2002database, bulygin2018, liang2021knowledge}, graph-based matching~\citep{drumm2007quickmig}, pre-training language model-based matching~\citep{gaowjw22, tu2023unicorn, zhang2023schema}, LLMs-based matching~\citep{zhang2024jellyfish}, and RAG-based matching~\citep{sheetrit2024rematch, seedat25bootstrapping, xu2024kcmf}. 
Although there are some recent works, such as ReMatch~\citep{sheetrit2024rematch} and Matchmaker~\citep{seedat25bootstrapping}, they leverage RAG over LLMs to address schema matching, but only the knowledge from the textual documents of the descriptions for the table and attribute is retrieved for augmentation.
KcMF~\citep{xu2024kcmf} retrieves knowledge from the self-built knowledge set, but the knowledge from the large commonsense and fact-based KGs in the real world has not been adapted and fully leveraged to augment the LLMs for schema matching. 
To some extent, the aforementioned methods achieve (semi-) automatic schema matching, but they are still incapable of resolving the semantic ambiguities and conflicts in some complex mapping cases because of the missing evidence and factual knowledge.

Besides schema matching and mapping, the field of data preparation/integration deals with several other critical challenges, e.g., data acquisition, extraction, cleaning, augmentation, transformation, and annotation; entity matching; ontology alignment; and data fusion, among others~\citep{DongR18,Chai0FL23,Berti-EquilleBM18}. Recent advances in deep learning and foundation models have also achieved promising results on many such tasks. Notable works include deep learning embedding-based data integration over relational databases~\citep{CappuzzoPT20,KoutrasFKL20}, and ontology matching~\citep{KolyvakisKK18}, pre-trained language models (PLMs) for entity matching~\citep{LLSDT20,Thirumuruganathan21}, and column type annotation~\citep{SuharaL0ZDCT22}, large language models (LLMs)-powered entity resolution~\citep{FanHFC00024,LiLHZSC24,PeetersB23} and other data wrangling tasks~\citep{NarayanCOR22,ZhangGCS24}, etc. 
The focus of our work is schema matching, which is different from the aforementioned problems. We do not consider cell values (nor rows/entities) in the table; instead, we directly aim at finding semantic correspondence among attribute names (i.e., column names) in tables from the source and target schema. Additionally, none of these works employ external and large-scale knowledge graph-based retrieval augmentation paradigms to enhance the LLM's performance in data integration tasks.

\noindent \textbf{Knowledge-augmented LLMs and RAG.} 
There are several ways in which external knowledge from KGs can be integrated with LLMs~\citep{MaCWKW25}. To leverage the strengths and capabilities of hierarchical and structured knowledge in complex reasoning, the KGs and text are aligned and interact to jointly train the language models~\citep{yasunaga2022deep, zhang2022greaselm} for knowledge-intensive tasks. 
In the phase of joint learning, the text and KG pair are aligned via entity linking and then fed into the cross-model encoder to bidirectionally fuse text and KG, thereby performing joint reasoning on text and KG. 
In addition to joint learning, knowledge‑based instruction tuning integrates graph representations into language generation by aligning LLM inference with the graph’s topological structure, enabling more effective use of graph knowledge for complex reasoning~\citep{buehler2025situ}. However, the resulting model must be re‑fine‑tuned whenever the KG is updated or when adapting to a new domain, because the method embeds only static KG information into the model parameters.

Graph reasoning approaches such as RoG~\citep{LuoLHP24} and ToG~\citep{SunXTW0GNSG24} guide LLM inference using KG-derived reasoning paths generated through BFS or beam search, rather than relying solely on static knowledge fusion. KG‑CoT~\citep{zhao2024kg} extends this idea by introducing chain‑of‑thought reasoning over graphs, producing explicit reasoning paths and confidence scores based on transition matrices. Agent‑based methods have also been explored, exemplified by KG‑Agent~\citep{jiang2025kg}, which treats the LLM as an agent that autonomously selects KG‑related tools and updates memory for iterative reasoning. Despite their strengths, these graph reasoning methods repeatedly search paths at each step of beam search or CoT, resulting in substantial computational overhead on large‑scale KGs.

Different from the aforementioned techniques, RAG aims to retrieve relevant knowledge from specific data sources, such as text chunks, tabular data, graph data, etc., to augment the capability and reliability of LLMs for knowledge-intensive NLP tasks~\citep{lewis2020retrieval}.
However, the mainstream RAG methods retrieve the relevant knowledge from the textual chunks, which ignore the structured information and inter-relations of these textual chunks.
To mitigate this limitation, Graph RAG~\citep{hu2024grag,xu2024retrieval} and KG-RAG~\citep{sanmartin2024kg} are proposed.
Instead of retrieving knowledge from textual chunks, Graph RAG directly retrieves the relevant paths from the graph data.
GraphRAG~\citep{hu2024grag} retrieves subgraphs from the textual graph that is indexed based on PLMs and then integrates the retrieved and pruned subgraphs with the query by aggregating and aligning the graph embeddings with text vectors based on GNNs.
Similarly, KG-based RAG aims to augment LLMs with KGs by leveraging the reasoning capability of knowledge from various knowledge sources for knowledge-intensive tasks. 
KG-RAG approaches retrieve the subgraph from the self-constructed KG based on unstructured text or textual chunks~\citep{sanmartin2024kg} or from the inherent KGs derived from the KGQA dataset~\citep{li2025simple} to augment the reasoning capability of LLMs for complex QA. 

Recent research works have demonstrated the remarkable performance of KG-RAG on several knowledge-intensive tasks such as text classification~\citep{wang2022entity}, sentiment analysis~\citep{li2023knowledge}, entity linking~\citep{shlyk-etal-2024-real}, KGQA~\citep{sanmartin2024kg, li2025simple, MaCWKW25}, recommendation~\citep{chen2022graph}, etc.
These mainstream Graph RAG and KG-RAG approaches usually retrieve knowledge from local and small-sized graphs built from text chunks or the inherent KGs derived from the KGQA dataset.
Our method differs from the existing KG-RAG on the KGQA; it leverages vector-based retrieval and graph traversal-based retrieval to retrieve the relevant subgraphs from large external KGs instead of constructed KGs based on textual chunks or the inherent KGs derived from the KGQA dataset. It mainly focuses on augmenting LLMs with KGs for complex schema matching tasks rather than the commonly investigated natural language processing and understanding tasks. 

\section{Conclusion}\label{sec6}
In this paper, we presented a novel knowledge graph-based retrieval-augmented generation for schema-matching solution, KG-RAG4SM, to address the limitations that traditional similarity-based and LLM-based schema-matching methods are incapable of resolving semantic ambiguities in complex mapping cases. 
Our framework differs from the emerging graph RAG approaches that retrieve knowledge from local and small-sized graphs built from text chunks. We proposed several subgraph retrieval methods from large KGs, followed by ranking-based subgraph refinement. Our empirical results confirmed that the vector-based KG triple retrieval and vector-based entity retrieval + BFS graph traversal methods are the optimal subgraph retrieval strategies.
The results of the extensive experimental evaluation demonstrated that our KG-RAG4SM outperforms PLMs and LLM-based SOTA methods in terms of effectiveness and efficiency. 
We investigated the scalability across KG domains from commonsense KG to domain-specific KG and the truthfulness of KG-RAG4SM with poorly relevant retrieved KG context.
Our case studies from real-world scenarios showed that the hallucination of LLMs for schema matching is well mitigated by our method. 

In this study, we augmented LLMs with relevant subgraphs retrieved from a large external KG via few-shot learning. 
However, resolving knowledge conflicts between the internal knowledge of LLMs and the retrieved KG,  examining retrieval performance across KGs varying in scale and connectivity, and developing more advanced graph traversal strategies to further optimize subgraph retrieval on large KGs remain interesting directions for future work. 
Additionally, we shall investigate how our KG-RAG paradigm can improve other LLM-based data integration and data fusion tasks.

\begin{acks}
Chuangtao Ma, Sriom Chakrabarti, and Arijit Khan acknowledge support from the Novo Nordisk Foundation grant (NNF22OC0072415). This work was also supported by the New National Excellence Program of the Ministry of Culture and Innovation Hungary (ÚNKP-22-4-I-ELTE-834).
\end{acks}

\bibliographystyle{ACM-Reference-Format}
\bibliography{sample-base}

\clearpage
\appendix

\section*{Appendix} \label{appendix}
This appendix provides prompts and templates (Appendix~\ref{app:prompt_templates}), detailed analysis (Appendix~\ref{app:detailed_analysis}), additional experiments and analysis (Appendix~\ref{app:additional_experiments_analysis}), and complexity and additional cost efficiency analysis (Appendix~\ref{app:complexity_additional_cost_efficiency_analysis}) beyond the main paper. 

\section{Prompt and Templates} \label{app:prompt_templates}

\subsection{Prompts for LLMs as Schema Matcher} \label{app:prompt_llms_as_schema_matcher}
\begin{tcolorbox}[boxrule=0.8pt, colback=gray!5!white, colframe=blue!25!white, coltitle=black, breakable]
\footnotesize
\noindent\textbf{Prompts for LLMs as Schema Matcher}

You are an expert in schema matching and data integration. 
Your task is to analyze attribute 1 with its textual description 1 and attribute 2 with its textual description 2 from the source and target schema in the given question, and specify if attribute 1 from the source schema is semantically matched with attribute 2 from the target schema.

---

Here are two examples of the schema matching questions with correct answers and explanations that you need to learn before you start to analyze the potential mappings:

\textbf{Example 1}:
Attribute 1 death-person\_id and its description 1 [d1]
Attribute 2 beneficiary\_summary-desynpuf\_id and its description 2 [d2]. 
\textit{Are attribute 1 and attribute 2 semantically matched with each other?}

Here is the \textbf{correct answer} and the explanations for the above given example question: \textbf{1}

\textbf{Explanation}: They are semantically matched with each other because both of them are unique identifiers for each person. Even if the death-person\_id refers to the unique identifier of the person in death table and beneficiary\_summary-desynpuf\_id refers to the unique identifier of the person beneficiary from beneficiary\_summary table, they are semantically matched with each other.
        
\textbf{Example 2:} ...

Remember the following tips when you are analyzing the potential mappings.
\textbf{Tips}:
(1) Some letters are extracted from the full names and merged into an abbreviation word.
(2) Schema information sometimes is also added as the prefix of abbreviation.
(3) Please consider the abbreviation case.  
          
---

Based on the above example and tips, please answer the following schema matching question:    

\textbf{Question:} \{\textbf{question}\}

Please respond only with the label: 1 If attribute 1 and attribute 2 are semantically matched with each other, otherwise respond with label: 0.
Do not mention that there is not enough information to decide.
\end{tcolorbox}

\subsection{Prompts for LLMs as Entity Retriever} \label{app:prompts_llms_as_entity_retriever}

\begin{tcolorbox}[boxrule=0.8pt, colback=gray!5!white, colframe=blue!25!white, coltitle=black, breakable]
\footnotesize
\textbf{Prompts for LLMs as Entity Retriever}

You are an expert in named entity recognition. Identify the list of relevant entities (preferably some entities corresponding to attribute 1 and its description, the others to attribute 2 and its description) in Wikidata that are helpful to answer the question from the following given text. 

Respond with only the list of entities with their values, separated by commas.

If you don't know the answer or if you can not find any entities in Wikidata, just say null. Do not make anything up.

---

For example:
Attribute 1 person-person\_id and its description 1 the person domain contains records that uniquely identify each patient in the source data who is time at-risk to have clinical observations recorded within the source systems; a unique identifier for each person. 
Attribute 2 beneficiary\_summary-desynpuf\_id and its description 2 beneficiary\_summary pertains to a synthetic medicare beneficiary; beneficiary code. \textit{Are attribute 1 and attribute 2 semantically matched with each other?}

The relevant entities in Wikidata are: \textbf{patient, unique identifier, beneficiary}. 

The response should be a list of entities in the given text, separated by commas, like this: patient, unique identifier, beneficiary.

---

\textbf{Question}: \textbf{\{question\}}
\end{tcolorbox}

\subsection{Prompts for LLMs as Subgraph Retriever} \label{app:prompts_llms_as_subgraph_retriever}

\begin{tcolorbox}[boxrule=0.8pt, colback=gray!5!white, colframe=blue!25!white, coltitle=black, breakable]
\footnotesize
\textbf{Prompts for LLMs as Subgraph Retriever}

You are an expert in relevant subgraph extraction from large Wikidata knowledge graphs. Identify the list of relevant subgraphs or paths in Wikidata that are helpful to answer the question from the following given text. Respond with only the list of entities with their values, separated by commas.
If you don't know the answer or if you can not find any relevant subgraph or path in Wikidata, just say null. Do not make anything up.

---

For example:

\textbf{Question}: Attribute 1 person-person\_id and its description 1 the person domain contains records that uniquely identify each patient in the source data who is time at-risk to have clinical observations recorded within the source systems; a unique identifier for each person. 
Attribute 2 beneficiary\_summary-desynpuf\_id and its description 2 pertains to a synthetic medicare beneficiary; beneficiary code. \textit{Are attribute 1 and attribute 2 semantically matched with each other?}

\textbf{Expected Response:} The relevant entities that are relevant to the above questions in Wikidata are: patient(Q181600), unique identifier (Q6545185), beneficiary (Q2596417). 
The response should be a list of paths or subgraphs between these entities in Wikidata, separated by "$|$", like this: patient (Q181600), subclass of (P279), customer (Q852835) $\rightarrow$ customer (Q852835), subclass of (P279), beneficiary (Q2596417).

Remember the following tips when you are analyzing and extracting subgraphs or paths from Wikidata knowledge graph.

\textbf{Tips:}
(1) Identify the relevant Wikidata entities that are relevant and helpful to answer the above questions, preferably some entities corresponding to attribute 1 and its description, and the others corresponding to attribute 2 and its description.
(2) Create the entity pair based on the identified entities from Wikidata knowledge graph.
(3) Search and find up to 3-hop paths or subgraphs between these entity pairs from Wikidata knowledge graph. 
(4) Please return the path in the format of "entity1\_value(entity1\_id), relation1\_value(relation1-property), entity2\_value(entity2\_id) $\rightarrow$ entity2\_value(entity2\_id), relation2\_value(relation2-property), entity3\_value(entity3\_id) " if the path is found.
(5) Please use the separator "$|$" to separate the path if multiple paths are found, like this: paths1 $|$ paths2 $|$ paths3.

---

\textbf{Question}: \textbf{\{question\}}
\end{tcolorbox}

\subsection{Prompts for Augmenting Generation with the Retrieved Subgraph} \label{app:prompts_augmenting_generation_with_retrieved_subgraph}

\begin{tcolorbox}[boxrule=0.8pt, colback=gray!5!white, colframe=blue!25!white, coltitle=black, breakable]
\footnotesize
\textbf{Prompts for Augmenting Generation with the Retrieved Subgraph}

You are an expert in schema matching and data integration. 
Your task is to analyze attribute 1 with its textual description 1 and attribute 2 with its textual description 2 from source and target schema in the given question, and specify if the attribute 1 from source schema is semantically matched with attribute 2 from the target schema.
\textbf{In some questions, there is the knowledge graph context that might be helpful for you to answer. In this case, you will need to consider the provided context to make the correct decision. }

---

Here are two examples of the question:

\textbf{Example 1}:
Attribute 1 death-person\_id and its description 1 [d1]
Attribute 2 beneficiarysummary-bene\_birth\_dt and its description 2 [d2] 
\textit{Are attribute 1 and attribute 2 semantically matched with each other?}

Here is the \textbf{knowledge graph context} that might be helpful for you to answer the above schema matching question: 
death (Q4), has part(s) of the class (P2670), date of death (Q18748141) $\rightarrow$  date of death (Q18748141), opposite of (P461), date of birth (Q2389905) $|$ human (Q5), has characteristic (P1552), age of a person (Q185836) $\rightarrow$ age of a person (Q185836), uses (P2283), date of birth (Q2389905)

Here is the \textbf{correct answer} and the explanations for the above given example question: 0

\textbf{Explanation}: They are not semantically matched with each other because death-person\_id is a unique identifier for each person in death table and bene\_birth\_dt is the date of birth of person in beneficiarysummary table. 

\textbf{From the above KG context, we can find that the date of death is the opposite of the date of birth, they are not semantically matched with each other.}

\textbf{Example 2}:...

Remember the following tips when you are analyzing the potential mappings.
\textbf{Tips}:(1) Some letters are extracted from the full names and merged into an abbreviation word. (2) Schema information sometimes is also added as the prefix of the abbreviation. (3) Please consider the abbreviation case. (4) \textbf{Please consider the knowledge graph context to make the correct decision when it is provided.}

---

Based on the provided example and the following knowledge graph context, please answer the following schema matching question: 

\textbf{\{question\}}

\textbf{KGs Context}: \textbf{\{paths\}} if paths and paths != ['null'] else "No available knowledge graph context, please make the decision on your own."

---

Please respond only with the label: 1 If attribute 1 and attribute 2 are semantically matched with each other, otherwise respond label: 0.
Do not mention that there is not enough information to decide.
\end{tcolorbox}

\subsection{Rule-based Instruction Tuning for Poorly-relevant Subgraph Exclusion} \label{app:rule-based_it_for_poorly-relevant_subgraph_exclusion}

\begin{tcolorbox}[boxrule=0.8pt, colback=gray!5!white, colframe=blue!25!white, coltitle=black, breakable]
\footnotesize
\textbf{Rule-based Instruction Tuning for Poorly-relevant Subgraph Exclusion}

You are an expert in schema matching and data integration. 

\textbf{(a) Your task is to analyze if attribute 1 with its textual description 1 and attribute 2 with its textual description 2 from the source and target schema are semantically matched with each other.}

\textit{In some questions, there is the knowledge graph context that might be helpful for you to answer. In this case, you will need to consider the provided context to make the correct decision.} 

\textbf{(b) Please give the relevance score between the given question and the provided knowledge graph context between 0 and 10 if the knowledge graph context is available.}
\textit{A score of 10 means the knowledge graph context is very relevant to the given question, and a score of 0 means the knowledge graph context is irrelevant to the given question. }

\textbf{(c) Please make the decision only based on your knowledge if the knowledge graph context relevance score is less than 7 or the provided knowledge graph context is unavailable.}

---

Here are some examples of the questions with knowledge graph context and their relevance score, reason for relevance score, correct answers, and explanations that you need to learn before you start to analyze the potential mappings:

\textbf{Example 1}:
Attribute 1 its description 1 [d1] 
Attribute 2 and its description 2 [d2] 
\textit{Are attribute 1 and atttribute 2 semantically matched with each other?}

Here is the \textbf{knowledge graph context} that might be helpful for you to answer the above schema matching question: 
...

\textit{Knowledge graph context relevance score}: ... 

\textit{Reason for relevance score}: ...

Here is the \textbf{correct answer} and the explanations for the above given example question: 0

\textit{Explanation}: ...

\textbf{Example 2}: ...

\textbf{Example 3}: ...

Remember the following tips when you are analyzing the potential mappings.
\textbf{Tips}: (1)...(2)...(3)... (4) \textbf{If conflicts exist between knowledge graph context and your own internal knowledge, please make the decision based on knowledge graph context.}

---

\textbf{(d) Based on the provided example, the following knowledge graph context, and relevance score, please answer the following schema matching question: }

\textbf{\{question\}}

\textbf{KG Context}: \textbf{\{paths\}} if paths and paths != ['null'] else "No available knowledge graph context, please make the decision on your own."

---

Please respond only with the label: 1 If attribute 1 and attribute 2 are semantically matched with each other, otherwise respond label: 0.
Do not mention that there is not enough information to decide.

\end{tcolorbox}

\subsection{Example of a Predefined SPARQL Template
for Remote BFS Retrieval}
\label{app:sparql_templates}
With a maximum search depth of three, we run separate predefined SPARQL templates to retrieve one-, two-, and three-hop paths between each entity pair. Rather than keeping only the shortest path, we combine all retrieved paths for ranking-based refinement, which preserves multiple semantic connections within the depth limit.
For example, the following SPARQL query retrieves three-hop paths between two Wikidata entities, unique key ($e_s$=Q3667566) and database ($e_e$=Q8513), and returns the intermediate entities and relations that connect them. 

\begin{tcolorbox}[
boxrule=0.8pt,
colback=gray!5!white,
colframe=blue!25!white,
coltitle=black,
breakable]
\scriptsize
\begin{verbatim}
PREFIX wd: <http://www.wikidata.org/entity/>

SELECT DISTINCT ?p1 ?mid1 ?p2 ?mid2 ?p3
WHERE {
  wd:Q3667566 ?p1 ?mid1 .
  ?mid1 ?p2 ?mid2 .
  ?mid2 ?p3 wd:Q8513 .

  FILTER(STRSTARTS(STR(?p1),
    "http://www.wikidata.org/prop/direct/"))
  FILTER(STRSTARTS(STR(?p2),
    "http://www.wikidata.org/prop/direct/"))
  FILTER(STRSTARTS(STR(?p3),
    "http://www.wikidata.org/prop/direct/"))
}
LIMIT 100
\end{verbatim}
\end{tcolorbox}

This example SPARQL query demonstrates how bounded-hop SPARQL queries identify semantic connections between entities for subsequent subgraph refinement.
The full SPARQL template additionally covers alternative edge orientations through \texttt{UNION} clauses.

\section{Detailed Analysis} \label{app:detailed_analysis}

\subsection{Detailed Analysis of Domain-specific UMLS \textit{vs.} Commonsense Wikidata KG}
We present examples of schema matching questions and their corresponding retrieved triples from UMLS KG and Wikidata KG in Table~\ref{tab-comparsion-kg-triple} and provide an additional comparison of KG triples.

\begin{table}[ht!]
\caption{The comparison between UMLS KG triples and Wikidata KG triples.}\label{tab-comparsion-kg-triple}
\begin{center}
\resizebox{\linewidth}{!}{
\begin{tabular}{p{5cm}|p{5cm}|p{5cm}|p{5cm}|p{5cm}} \toprule
Question & UMLS KG Triple & Wikidata KG Triple & Comparison    &Role of KG Triple\\ \midrule

Attribute 1 \texttt{drug\_exposure-} \texttt{provider\_id} and its description 1 [d1] Attribute 2 \texttt{medications-code} and its description 2 [d2]. \textit {Are attribute 1 and attribute 2 semantically matched with each other?} & \texttt{Medication code:ID:Pt:$\hat{}$Patient:Nom, has scale, Nom $|$ Medication code:ID:Pt:$\hat{}$Patient:Nom, has time aspect, Point in time (spot)} & \texttt{DrugBank (Q1122544), instance of (P31), online database (Q7094076) $|$ PKR inhibitor (Q5007568), instance of (P31), chemical compound (Q11173)} & The UMLS KG triple defines a "Medication code" as an identifier of patient that is recorded at a specific point, while Wikidata KG triples describe external entities of drugbank, database, inhibitor, and a chemical compound, both of which are not useful for the given question, as they do not provides any information of provider and medication. & The UMLS KG triple provides a connection between medication code and identifier, which serves as a useful background knowledge for decision-making. \\ \hline
Attribute 1 \texttt{drug\_exposure-} \texttt{stop\_reason} and its description 1 [d1] Attribute 2 \texttt{medications-reason- description} and its description 2 [d2]. \textit{Are attribute 1 and attribute 2 semantically matched with each other?}
& \texttt{Drug Exposure (C41362): Contact with drug, isa, Chemical Exposure (C36290)} $|$ \texttt{reason for stopping medication (120234), isa, Medication discontinued (274512008)} $\rightarrow$ \texttt{has\_associated\_procedure, Drug therapy (416608005), has\_direct\_substance, Drug or medicament (410942007)}.
 & \texttt{Parenteral Drug Association (Q7136792) $\rightarrow$ instance of(P31) $\rightarrow$ nonprofit organization (Q163740) $|$ PKR inhibitor (Q5007568) $\rightarrow$ instance of (P31) $\rightarrow$ chemical compound (Q11173)} &  The UMLS KG triple provides direct semantic concepts related to the action of stopping a drug and the context of drug therapy. The Wikidata triples, conversely, refer to a non-profit organization (The Parenteral Drug Association)  and define a specific class of chemical compound, which is not relevant to the given schema matching question.  &The UMLS KG triple conveys a connection between the drug exposure stop
reason and the reason for stopping medication, but this connection is a sub-concept relation, which might mislead LLMs to generate a false positive answer.\\ \hline
Attribute 1 \texttt{device\_exposure-} \texttt{device\_type\_concept\_id} and its description 1 [d1] Attribute 2 \texttt{providers-id} and its description 2 [d2]. \textit{Are attribute 1 and attribute 2 semantically matched with each other?} & \texttt{Provide guidance on organization of airy, light, and clean environment, CHD, Discharge Planning: Home Preparation $|$ Exclusive Provider Organization, RO, Exclusive Provider Organization Plan} &\texttt{Electronic Product Code (Q1327374), instance of(P31), unique identifier (Q6545185) $|$ ATNAA (Q4654746), instance of (P31), medical device (Q6554101)} & The UMLS KG triple focuses on organization, discharge, provider terms rather than schema-level metadata, while Wikidata KG triple provides a direct relation between Identifier and medical device. & Wikidata KG triple provides a direct relation between the identifier and medical device that helps to understand the connection between device ID and product/device.\\ \bottomrule
\end{tabular}}
\end{center}
\end{table}
We observed that the UMLS KG triple provides more healthcare-technical specialized knowledge than the Wikidata KG, but it is not always useful for metadata-level schema matching. 
Although all of our schema matching datasets originate from healthcare databases, the questions of schema-matching pairs including table name (\texttt{admission, patient, visit, insurance, service}, etc.), attribute name (\texttt{admission\_location, patient\_weight, visit\_date, insurance\_id, service\_provider}, etc.) from source and target data model; are more relevant to the entities and taxonomic relations (e.g., \texttt{patient, instance of, human}, etc.) from commonsense KGs rather than medical entities and relations (e.g, \texttt{diabetes status:Prid: Pt: $\hat{}$ Patient:Nom, has\_supersystem, patient}, etc.) from domain-specific UMLS KGs.

To further investigate this interesting finding, we manually selected 160 schema questions that cover more professional terms (e.g., \texttt{drug\_exposure}, \texttt{medications}, \texttt{immunization}, etc.) in the healthcare domain from the \textsf{Synthea} dataset and compared the results of KG-RAG4SM with Jellyfish-8B and the top-2 ranked triples from UMLS and Wikidata KGs in Table~\ref{tab:selected-questions-umls-wikidata}.
\begin{table}[ht!]
\caption{KG-RAG4SM with Jellyfish-8B and UMLS KGs on selected 160 questions from \textsf{Synthea} dataset.} \label{tab:selected-questions-umls-wikidata}
\small
\begin{center}
\begin{tabular}{c|c|c}
\toprule
\multirow{2}{*}{Metric (\%)} & UMLS KG Triple & Wikidata KG Triple \\ 
\cline{2-3} & Top-2 Ranked & Top-2 Ranked \\ 
\midrule
\texttt{P}               & 100& 100\\ 
\hline
\texttt{R}               & 27.27& 18.18\\ 
\hline
\texttt{F1}             & 42.86& 30.77\\ 
\bottomrule
\end{tabular}
\end{center}
\end{table}
The results demonstrate that KG-RAG4SM with a domain-specific UMLS KG triple outperforms the approach with a Wikidata KG triple across metrics on the selected 160 schema-matching questions containing highly healthcare-specialized terms.
This confirms that a domain-specific KG works better than a commonsense KG in KG-RAG4SM in the scenario of a schema matching question that contains several healthcare domain-specialized terms.  

\subsection{Detailed Analysis of Baselines with a 0\% or 100\% Score}

\begin{figure}[ht!]
  \centering
  \begin{subfigure}[t]{0.35\linewidth}
    \centering
    \includegraphics[width=0.96\linewidth]{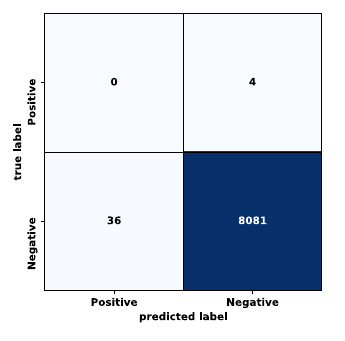}
    \caption{Jellyfish-8B baseline on \textsf{EMED}}
    \label{fig:cm_jellyfish}
  \end{subfigure}\hspace{1em} 
  \begin{subfigure}[t]{0.35\linewidth}
    \centering
    \includegraphics[width=\linewidth]{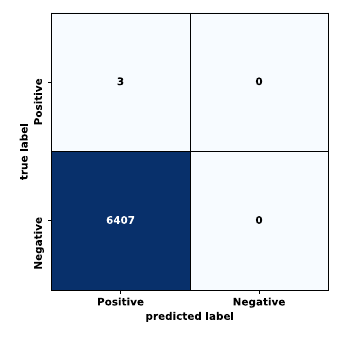}
    \caption{SMAT baseline on \textsf{MIMIC}}
    \label{fig:mc-mimic}
  \end{subfigure}

  \begin{subfigure}[t]{0.35\linewidth}
    \centering
    \includegraphics[width=\linewidth]{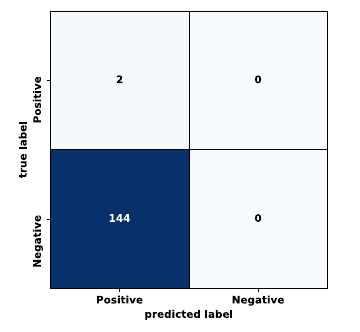}
    \caption{SMAT baseline on \textsf{Bank}}
    \label{fig:cm-bank}
  \end{subfigure}\hspace{1em} 
  \begin{subfigure}[t]{0.35\linewidth}
    \centering
    \includegraphics[width=0.94\linewidth]{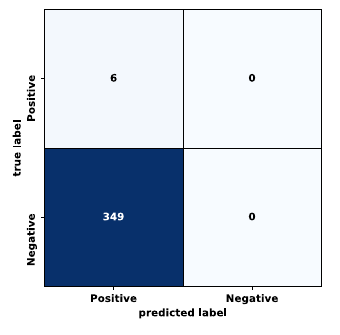}
    \caption{SMAT baseline on \textsf{IMSA}}
    \label{fig:cm-imsa}
  \end{subfigure}

  \caption{Confusion matrix of Jellyfish-8B and SMAT baselines.}
  \label{fig:cm-smat}
\end{figure}

\noindent \textbf{Detailed Analysis of Jellyfish with a 0\% Score}
We present detailed results and a discussion of the Jellyfish-8B baseline, as we observed that Jellyfish-8B achieved a 0\% score on the \textsf{EMED} dataset.
Figure~\ref{fig:cm_jellyfish} shows the confusion matrix of Jellyfish-8B on the \textsf{EMED} baseline.

\captionsetup[subfigure]{skip=-2pt, font=small}

The reasons behind this are twofold: \textbf{(1)} the dataset is unseen, and \textbf{(2)} it is highly imbalanced. Jellyfish-8b is a fine-tuned model based on Meta-Llama-3-8B-Instruct with several data preprocessing datasets, including schema matching datasets, i.e., \textsc{MIMIC} and \textsc{Synthea}. 
The observed 0\% score arises from the fact that the fine‑tuned Jellyfish‑8B model fails to identify any true positive matches (\texttt{TP}=0) on the \textsc{EMED} dataset, which is newly created and unseen for this model. In addition, \textsc{EMED} is highly imbalanced and negative‑dominant, with very few positive matching pairs in the ground truth. This further highlights the limited robustness of the fine‑tuned model when applied to an unseen and imbalanced dataset.

\noindent \textbf{Detailed Analysis of SMAT Baseline with 100\% Recall}
We present detailed results and a discussion of the SMAT baseline, as we observed that SMAT achieved 100\% recall across all datasets.
Figure~\ref{fig:cm-smat} shows the confusion matrix of the SMAT baseline, which demonstrates that SMAT tends to return a positive answer and results in 100\% recall. 
            
The confusion matrix shows that SMAT exhibits a strong bias toward predicting the positive class, assigning positive labels to all matching pairs while failing to identify any negatives. This behavior indicates clear overfitting, as the model does not learn the discriminative features needed to separate positive from negative cases. Moreover, the dataset is highly imbalanced and negative‑dominant, with only a small fraction of positive pairs in the ground truth. As a result, SMAT aggressively classifies negative samples as positive, capturing the few true positives but at the cost of substantial misclassification. In contrast, our proposed method achieves a more balanced trade‑off between recall and precision, leading to more robust schema‑matching performance for data integration.

\section{Additional Experiments and Analysis} \label{app:additional_experiments_analysis}
\subsection{Ablation Study on the Impact of Embedding Method}
To assess the effect of the embedding method on vector‑based KG triple retrieval in KG‑RAG4SM with Wikidata5M KG, we conduct an ablation study comparing our 768‑dimensional RoBERTa‑base embeddings with \texttt{jina‑embeddings‑v3}, a state‑of‑the‑art multilingual model with 512 dimensions. 
Table~\ref{tab:ablation-study-embed-method} summarizes the performance of KG-RAG4SM with different embedding models and GPT-4o-mini on the \textsf{CMS} dataset.

\begin{table}[ht!]
\caption{Impact of the embedding method in KG-RAG4SM with GPT-4o-mini on \textsf{CMS} dataset.} 
\label{tab:ablation-study-embed-method}
\small
\begin{center}
\begin{tabular}{c|c|c||c|c}
\toprule
\multirow{2}{*}{Metric (\%)} & \multicolumn{2}{c||}{KG-RAG4SM (RoBERTa-base)} & \multicolumn{2}{c}{KG-RAG4SM (jina-embeddings-v3)} \\ 
\cline{2-5} 
                                             & Top-1 Ranked& Top-2 Ranked& Top-1 Ranked& Top-2 Ranked\\ 
\midrule
\texttt{P}               & \textbf{62.50}& \underline{52.38}& 44.44& 33.33\\ 
\hline
\texttt{R}               & \underline{40.00}& \textbf{44.00}& 32.00& 28.00\\ 
\hline
\texttt{F1}             & \textbf{48.70}& \underline{47.82}& 37.21& 30.43\\ 
\bottomrule
\end{tabular}
\end{center}
\end{table}

The results demonstrate that KG-RAG4SM with RoBERTa-base consistently outperforms the version with \texttt{jina-\allowbreak embeddings-v3}, which indicates the strength of a high-dimensional embedding method in our KG-RAG4SM. 
This suggests that higher‑dimensional embeddings capture richer semantic information for vector retrieval, albeit at increased indexing and search cost. 
This also highlights that our RoBERTa-base is a trade-off configuration to balance the performance and the computational cost of indexing and vector search for vector-based retrieval in KG-RAG4SM.

\subsection{Detailed Experiment and Analysis on BFS Search Depth} 
\noindent \textbf{Sensitivity Analysis on BFS Search Depth.}
To analyze the sensitivity of BFS search depth in KG-RAG4SM, we conduct a new empirical study on the sensitivity of BFS search depth with settings of 2, 3, and 4 in KG-RAG4SM with GPT-4o-mini and Wikidata5M KG on the \textsf{CMS} dataset. 
Table~\ref{tab:bfs-depth-sensitivity-analysis} shows the results of KG-RAG4SM with different BFS search depths.

\begin{table}[ht!]
\caption{Sensitivity analysis of BFS search depth in KG-RAG4SM with GPT-4o-mini on \textsf{CMS} dataset.} \label{tab:bfs-depth-sensitivity-analysis}
\begin{center}
\small
\begin{tabular}{c|c|c|c}
\toprule
\multirow{2}{*}{Metric (\%)} & 2& 3& 4\\ 
\cline{2-4} & Top-1 Ranked & Top-1 Ranked & Top-1 Ranked \\ 
\midrule
\texttt{P}               & \textbf{46.67}& 42.11& \underline{44.44}\\ 
\hline
\texttt{R}               & \underline{28.00}& \textbf{32.00}& \textbf{32.00}\\ 
\hline
\texttt{F1}             & 35.00& \underline{36.36}& \textbf{37.21}\\ 
\bottomrule
\end{tabular}
\end{center}
\end{table}

The results show that KG-RAG4SM with a BFS search depth of 4 achieves the best \texttt{F1} score, while BFS depth 3 yields comparable results to KG-RAG4SM with BFS depth 2.  
This highlights a trend that increasing the depth from 2 to 3 and 4 results in an increase in the \texttt{F1} score.
However, the design choice of BFS search depth in \textsf{KG-RAG4SM} is determined by jointly considering overall matching performance (\texttt{F1} score), retrieval time, token cost, and end-to-end latency.
We justify our default design choice of BFS search depth of 3 as a favorable accuracy–efficiency trade-off in the following BFS search time evaluation.

\noindent \textbf{Remote BFS vs. BFS in Local Memory.}
We further evaluate KG‑RAG4SM by comparing the number of retrieved paths and triples, along with the average BFS search time per question, across search depths of 2 to 4. The analysis considers both remote BFS via API and local BFS in‑memory settings.
Specifically, remote BFS search uses the full Wikidata KG through its SPARQL endpoint, whereas local in-memory BFS operates over the Wikidata5M. Since Wikidata5M contains only a subset of the entities and relations in Wikidata, the two configurations do not necessarily provide the same graph connectivity or retrieved paths, even when using the same BFS depth. Consequently, the two configurations may retrieve different numbers of paths and triples.
For the local in-memory implementation, the per-question BFS search times reported in Table~\ref{tab:bfs-depth-sensitivity-analysis-time} are distinct from the one-time graph preparation and process-initialization costs. The serialized Wikidata5M adjacency representation and entity/relation labels require 1.23~GB of local storage, while fresh-process initialization takes 57.90~s on average over three runs, comprising 49.59~s for loading the adjacency graph and 8.32~s for loading the labels. Peak resident memory during initialization reaches 8.31~GB. We report the complete deployment footprint and initialization breakdown in Appendix~\ref{sec:graph-retrieval-deployment-efficiency}.

\begin{table}[ht!]
\caption{Comparison of the number of retrieved paths, triples, and average BFS search time per question with remote and local BFS implementations varying in search depth.}
\label{tab:bfs-depth-sensitivity-analysis-time}
\small
\begin{center}
\begin{tabular}{c|c|c|c||c|c|c}
\toprule
 & \multicolumn{3}{c||}{Remote BFS via API }& \multicolumn{3}{c}{BFS in Local Memory}\\ \hline
Depth & 2& 3& 4 & 2& 3&4\\ 
\midrule
\#Path & 3.25& 4.11& 17.83 & 1.47& 3.66&13.34\\ 
\hline
\#Triple & 6.29& 8.88& 63.76 & 2.60& 8.79&48.07\\ 
\hline
Time (Sec)& 39.09& 83.77& 98.45 & 0.001& 5.47&11.80\\ 
\bottomrule
\end{tabular}
\end{center}
\end{table}

Table \ref{tab:bfs-depth-sensitivity-analysis-time} shows that increasing the BFS search depth leads to a steep growth in retrieved triples and paths, as well as average BFS search time per question, both in remote BFS search via API and local BFS search in memory. 
As shown in Table~\ref{tab:bfs-depth-sensitivity-analysis-time}, at a BFS depth of three, remote retrieval produces an average of 4.11 paths and 8.88 triples per question, compared with 3.66 paths and 8.79 triples using local Wikidata5M. The difference becomes more pronounced at depth four, where remote retrieval obtains 17.83 paths and 63.76 triples, compared with 13.34 paths and 48.07 triples locally. These observations are consistent with the broader graph connectivity available through full-scale Wikidata, which can expose additional intermediate entities and semantic relationships not represented in Wikidata5M.
These results also demonstrate that a BFS depth of 3 offers a favorable accuracy--efficiency trade-off for KG-RAG4SM in both remote API-based and local in-memory search settings. It provides richer retrieval, covering substantially more multi-hop paths than depth 2, while maintaining lower search time and end-to-end latency, and avoiding the excessive delays and timeout risks observed at depth 4. This trade-off between traversal depth, retrieved knowledge coverage, and search latency is a fundamental consideration when deploying BFS-based retrieval over large-scale knowledge graphs such as Wikidata5M, as broader coverage often introduces noisy information that degrades overall performance.

The results also demonstrate that KG-RAG4SM with local BFS search in memory is always faster than remote BFS search via the Wikidata API, which further confirms the deployment flexibility of KG-RAG4SM with a local in-memory BFS strategy in real-world data integration scenarios.

\subsection{Sensitivity Analysis of Relevance Score Threshold in Instruction Tuning-based Knowledge Exclusion} 
To analyze the sensitivity of the relevance score threshold in instruction tuning-based knowledge exclusion, we conduct a new empirical study on the relevance score threshold with settings of 3, 5, 7, and 9 in KG-RAG4SM with Jellyfish-8B and UMLS KGs on the \textsf{Synthea} dataset. 
Table~\ref{tab:relevance-scoring-sensitivity-analysis} shows the results of KG-RAG4SM with different settings of the relevance score threshold in instruction tuning-based knowledge exclusion.

\begin{table}[ht!]
\caption{Sensitivity analysis of relevance score threshold in IT-based knowledge exclusion of KG-RAG4SM with Jellyfish-8B and UMLS KGs on the \textsf{Synthea} dataset.} \label{tab:relevance-scoring-sensitivity-analysis}
\small
\begin{center}
\begin{tabular}{c|c|c|c|c}
\toprule
\multirow{2}{*}{Metric (\%)} & 3& 5& 7& 9\\ 
\cline{2-5} & Top-2 Ranked & Top-2 Ranked & Top-2 Ranked & Top-2 Ranked \\ 
\midrule
\texttt{P}               & 21.43& \underline{23.08}& \textbf{33.33}& 21.43\\ 
\hline
\texttt{R}               & 27.27& 27.27& \textbf{45.45}& 27.27\\ 
\hline
\texttt{F1}             & 24.00& \underline{25.00}& \textbf{38.46}& 24.00\\ 
\bottomrule
\end{tabular}
\end{center}
\end{table}

We observed that the \texttt{P} and \texttt{F1} increase as the relevance score threshold rises from 3 to 7, but performance declines when the threshold is further increased from 7 to 9. 
This is mainly because increasing the relevance threshold from 3 to 7 filters out less relevant or noisy knowledge and improves performance, but raising it further to 9 becomes too strict for knowledge exclusion.
This confirms the effectiveness of our designed instruction-tuning-based knowledge exclusion in excluding weakly relevant or noisy knowledge without discarding relevant and useful knowledge for reasoning.  
These results further indicate that a threshold of 7 provides a balance between retaining relevant knowledge and excluding noise. 
More broadly, the analysis demonstrates that the relevance threshold is a key parameter in instruction tuning-based knowledge exclusion, enabling effective filtering of irrelevant triples when set appropriately.

\subsection{LLM-based Subgraph Retrieval} \label{sec:llm-subgraph retrieval}

In LLM-based KG retrieval, we treat the LLM as a natural language interface for the commonsense KG. 
This is mainly based on the fact that commonsense KGs, such as Wikidata, are well known to LLMs, as the commonsense KGs or their open textual data forms might have already been utilized to pre-train LLMs.
Accordingly, we designed an LLM-based entity retrieval + BFS and an LLM-based subgraph retrieval. 

In the LLM-based entity retrieval + BFS, the interaction pipeline is summarized as: \textit{(1) Entity/Relation ID Generation:} The LLM is prompted to output the Wikidata entity identifier (QID) and relation identifier (PID) that are relevant to the given schema matching question in context learning; \textit{(2) ID fetcher and Verification:} We verify and fetch the name of the entity and the relation with the identifier from Wikidata via the Wikidata REST API; \textit{(3) Subgraph Search:} We execute the subgraph search via BFS with the verified entity and relation.

In LLM-based subgraph retrieval, the interaction pipeline is summarized as: \textit{(1) Path Generation:} The LLM is prompted to output the paths from Wikidata that are relevant to the given schema-matching question; \textit{(2) ID fetcher and Verification:} We verify the paths by fetching the name of the entity and relation with identifiers from Wikidata via the Wikidata REST API.

Our empirical evaluation (\textbf{Abl-1: subgraph retrieval}) results reveal that the LLM-based subgraph retrieval method works on commonsense KGs because LLMs know the concepts and relations of Wikidata, but sometimes it returns the entity and relation identifiers that actually do not exist in Wikidata. 
However, LLM-based subgraph retrieval works for commonsense KGs, whereas it may not work as effectively for domain-specific KGs, as LLMs are not familiar with them.

\subsection{KG-RAG4SM with Domain KGs \textit{vs.} LLM-based Schema Matching}
To highlight the contribution of retrieved external knowledge rather than the LLM’s internal parametric knowledge, we compare KG‑RAG4SM using the domain‑specific UMLS KG against LLM‑based schema matching with the same backbone models under a low‑overlap setting (Table \ref{tab:comparsion-umls-kg-rag-llm-based-em}). 
\begin{table}[ht!]
\caption{KG-RAG4SM with Jellyfish-8B and UMLS KGs vs.\ LLM-based SM on \textsf{Synthea} dataset.} \label{tab:comparsion-umls-kg-rag-llm-based-em}
\small
\begin{center}
\begin{tabular}{c|c|c|c}
\toprule
\multirow{2}{*}{Metric (\%)} & \multicolumn{2}{c|}{KG-RAG4SM With Rule-based IT} & \multirow{2}{*}{LLM-based SM}\\ 
\cline{2-3} 
                                             & Top-1 Ranked &Top-2 Ranked  & \\ 
\midrule
\texttt{P}               & \textbf{35.71}&\underline{33.33} & 10.52\\ 
\hline
\texttt{R}               & \textbf{45.45}&\textbf{45.45} & 36.36\\ 
\hline
\texttt{F1}             & \textbf{40.00}&\underline{38.46} & 16.32\\ 
\bottomrule
\end{tabular}
\end{center}
\end{table}
The results show that KG‑RAG4SM with UMLS consistently outperforms the LLM‑only baseline across all metrics. This confirms that the performance gains stem from the retrieved external knowledge, rule‑based instruction tuning, and structural grounding, rather than from the LLM’s inherent parameter knowledge.

\subsection{Qualitative Analysis on Improvement} \label{sec:qualitative-analysis-improvement}
\begin{figure}[ht!]
\centering
\includegraphics[width=0.7 \textwidth]{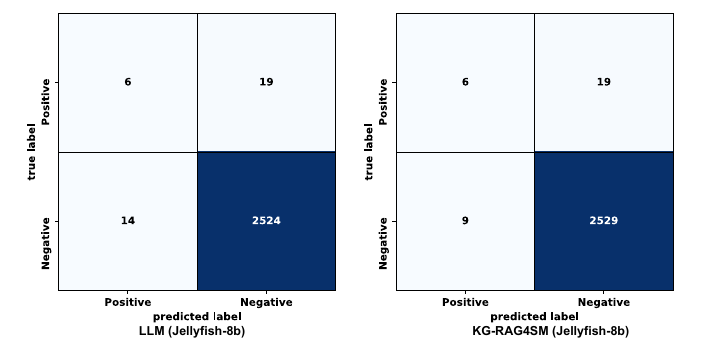}
\caption{Confusion matrix of LLM-based schema matching and KG-RAG4SM on \textsf{CMS} dataset. } \label{fig:cm_comparsion_jellyfish}
\end{figure}

To qualitatively illustrate the impact of retrieved subgraphs in KG‑RAG4SM, we present the confusion matrices (\texttt{TP}, \texttt{FN}, \texttt{FP}, \texttt{TN}) for KG‑RAG4SM using Jellyfish‑8B as the backbone LLM with top‑1 triples retrieved via vector‑based entity retrieval plus BFS search, compared with the LLM baseline on the \textsf{CMS} dataset. As shown in Figure \ref{fig:cm_comparsion_jellyfish}, KG‑RAG4SM achieves 2,529 \texttt{TN} and 9 \texttt{FP}, whereas the LLM‑only baseline yields 2,524 \texttt{TN} and 14 \texttt{FP}.
This demonstrates that incorporating the retrieved subgraph increases \texttt{TN} and reduces \texttt{FP} without decreasing \texttt{TP} or increasing \texttt{FN}. The reduction in \texttt{FP} indicates that the external KG subgraph provides structural factual context that enables more reliable reasoning, allowing KG‑RAG4SM to generate correct answers on complex schema‑matching cases.
These findings are further supported by the case study in \S\ref{susec:case_study}, which shows that the structural knowledge contained in the retrieved subgraph enhances the LLM’s reasoning ability in complex schema‑matching scenarios and reduces hallucinations during matching.

\section{Complexity and the Additional Cost Efficiency Analysis} \label{app:complexity_additional_cost_efficiency_analysis}

\subsection{Complexity Analysis and Empirical Verification} \label{sec:compexity-scalability-analysis}
We provide a time complexity analysis of vector retrieval and BFS graph traversal, along with empirical verification of the computational feasibility of KG-RAG4SM on large-scale KGs.

\noindent \textbf{Time Complexity Analysis.}
Let $G=(V, E)$ denote a KG, where $V$ is a set of entities with $n=|V|$ and $E$ is a set of edges with $m=|E|$. The average branching factor is the average out-degree of a node in the KG: $b = m/n$. 
For a given query $q$, the complexity is derived as follows:
\textbf{(1)} \textit{Vector Retrieval:} For cosine similarity search over KG embeddings in a dimension $d$ vector space, the time complexity scales linearly with the KG size $n$: $O(n \cdot d)$. 
In practice, Approximate Nearest Neighbor (ANN) indexes such as HNSW can reduce query cost substantially, often exhibiting sublinear empirical search behavior: $O(\log n \cdot d)$, although worst-case complexity depends on the index and data distribution.
\textbf{(2)} \textit{BFS Traversal:} Unlike standard BFS that may visit all nodes $O(|V| + |E|)$ for each entity, our designed BFS search is constrained by the number of entity pairs formed based on the number of seed entities $V_S$ in the initial vector-based entity retrieval and the maximum BFS search depth $D_{max}$.
The BFS search executes for each entity pair, so the total time complexity for the graph traversal phase in the worst case expands exponentially with depth $D_{max}$ is: $ O(|V_S|^2 \cdot b^{D_{max}})$. 
Therefore, the time complexity of the subgraph retrieval is $O(\log n \cdot d + |V_S|^2 \cdot b^{D_{max}})$, which is dominated by the maximum BFS search depth $D_{max}$ and the local connectivity of the KG, and is not directly determined by the overall number of entities in the KG.

\noindent \textbf{Empirical Analysis and Feasibility Verification.}
Since the above complexity analysis indicates that the time complexity of KG retrieval in \textsf{KG-RAG4SM} depends primarily on $D_{max}$ and the local connectivity of the KG rather than the KG size, our empirical analysis focuses on the BFS traversal depth ($D_{max}$) instead of the KG scale.
To evaluate the impact of BFS depth ($D_{max}$), we compare remote traversal over full-scale Wikidata with local in-memory traversal over Wikidata5M ($n \approx 4.6M$, $m \approx 20.6M$).
Table~\ref{tab:bfs-depth-sensitivity-analysis-time} reports the number of retrieved paths and triples, alongside the average search time per question, as the BFS search depth $D_{max}$ varies from 2 to 4.
Specifically, as shown in Table~\ref{tab:bfs-depth-sensitivity-analysis-time}, increasing the depth for remote full-Wikidata BFS, from 3 to 4, results in a $7\times$ increase in retrieved triples and causes the average search time to reach 98.45 seconds per query.
These results indicate that increasing the BFS search depth $D_{max}$ significantly increases the number of retrieved triples, but also incurs higher BFS search latency and a greater risk of API timeouts.
This suggests that a trade-off between BFS search depth and search time is necessary to balance the number of retrieved triples with search time when adapting to large-scale KGs.

\subsection{Graph Retrieval and Deployment Efficiency} \label{sec:graph-retrieval-deployment-efficiency}

\noindent \textbf{Retrieval Latency and Runtime Analysis.}
To further evaluate the efficiency and feasibility of KG-RAG4SM on a large-scale KG, we analyze the vector search runtime and BFS retrieval latency of our KG-RAG4SM.
We also provide a BFS search latency optimization strategy for practical system deployment of KG-RAG4SM by extending BFS from remote search via Wikidata endpoint (full Wikidata KG) access to local BFS search in memory (Wikidata5M). 
We measure the runtime of vector search and BFS graph traversal, both in remote search and local memory search, on the \textsf{CMS} dataset, and then report the average retrieval latency per question. The latency is divided into the initial vector search and the graph traversal.

\begin{table}[ht!]
\centering
\caption{End-to-end latency of vector-based entity retrieval + BFS subgraph retrieval on \textsf{CMS} dataset.}\label{tab:runtime-latency}
\small
\begin{tabular}{@{}l|r|r}
\toprule
Latency Breakdown (Sec) &Remote BFS via API &BFS in Local Memory\\ \midrule
Retrieval Latency & 83.77&5.47\\ \hline
  \hspace*{3mm} - Vector Search    & 0.0001                    &0.0001                    \\ \hline
  \hspace*{3mm} - BFS Graph Traversal    & 83.7729&5.4690\\ 
 \bottomrule
\end{tabular}
\end{table}

As summarized in Table~\ref{tab:runtime-latency}, the vector search stage averages approximately 0.0001 seconds per question.
The BFS graph traversal stage is the most time-consuming phase, as it involves retrieving relational paths from the large-scale KG. The average latency per question is 83.77 s and 5.47 s, under the settings of remote BFS search over full-scale Wikidata through its SPARQL endpoint and local in-memory BFS search over Wikidata5M, respectively, with maximum BFS search depth $D_{max}=3$.
Specifically, KG-RAG4SM with in-memory BFS search achieves approximately a $15\times$ reduction in retrieval latency compared to remote BFS search over full-scale Wikidata. This is mainly because remote BFS involves network communication and query processing through the Wikidata SPARQL endpoint, whereas local BFS directly traverses the in-memory Wikidata5M graph.
Moreover, since Wikidata5M contains only a subset of the entities and relations in full-scale Wikidata, differences in graph coverage may also affect the number of retrieved paths and triples. Therefore, the observed latency difference reflects both the retrieval implementations and the underlying graph configurations, rather than a comparison over identical graphs.
These results demonstrate that the local in-memory BFS strategy is computationally efficient and deployable under realistic implementation settings.

We also observe a significant trend: as the search depth increases, the number of retrieved triples grows rapidly, directly impacting retrieval and graph traversal cost.
Based on the sensitivity analysis results in Table~\ref{tab:bfs-depth-sensitivity-analysis} and Table~\ref{tab:bfs-depth-sensitivity-analysis-time}, a BFS search depth of $D_{max}=3$ offers a favorable balance between factual grounding and computational efficiency, providing sufficient contextual factual knowledge to ground the LLM's inference without incurring prohibitive latency or timeout risks.
Although the end-to-end retrieval latency is dominated by the graph traversal stage, our theoretical and empirical analysis demonstrates the practical feasibility and computational efficiency of \textsf{KG-RAG4SM} using Wikidata5M for local BFS and full-scale Wikidata for remote BFS, with a balanced BFS search depth.

\noindent \textbf{Local Deployment and Initialization Cost.}
To further assess the practical deployment requirements of the local in-memory BFS implementation, we report its storage footprint, graph loading and construction cost, and hardware requirements in Table~\ref{tab:local-bfs-deployment-cost}.

\begin{table}[ht!]
\centering
\caption{Deployment cost and hardware requirements of the local in-memory BFS implementation on Wikidata5M.}
\label{tab:local-bfs-deployment-cost}
\small
\resizebox{\linewidth}{!}{
\begin{tabular}{l|l}
\toprule
Deployment Aspect & Hardware Requirements and Result \\ 
\midrule

Local storage footprint 
& 1.23 GB (0.72 GB adjacency cache + 0.51 GB entity/relation labels) \\ \hline

Graph + labels memory footprint 
& 8.29 GB RSS \\ \hline

Peak initialization memory 
& 8.31 GB RSS \\ \hline

One-time adjacency-cache construction 
& 111.97 s \\ \hline

Fresh-process graph loading time 
& 49.59 s (average over 3 runs) \\ \hline

Separate graph index 
& None; BFS directly uses the serialized bidirectional adjacency structure \\ \hline

Measurement hardware 
& Ubuntu 22.04.5, OpenStack KVM, 32$\times$ Intel Xeon Skylake vCPUs, 64 GB RAM \\ \hline

GPU requirement 
& None; local BFS operates entirely in CPU memory \\ 

\bottomrule
\end{tabular}}\end{table}

The local Wikidata5M implementation requires 1.23~GB of persistent storage and 8.29~GB of resident memory, with peak initialization memory usage reaching 8.31~GB. Constructing the serialized adjacency cache is a one-time preprocessing operation requiring 111.97~s, while loading the graph in a fresh process takes 49.59~s on average over three runs. No separate graph index or GPU is employed for local BFS traversal.

Overall, by implementing BFS search in local memory, KG-RAG4SM achieves acceptable latency for real-world applications, highlighting its deployment flexibility in schema matching.
However, a systematic empirical evaluation of subgraph retrieval across KGs varying in scale and topological density is beyond the scope of this work.
Developing more advanced graph traversal strategies to further optimize subgraph retrieval on large KGs remains important for future investigation.
\end{document}